\documentclass[aps,nofootinbib,nobibnotes,noeprint,preprintnumbers,floatfix]{revtex4-2}
\usepackage{graphicx} 
\usepackage{hyperref}
\usepackage[utf8]{inputenc}
\usepackage{graphicx}
\usepackage{amsmath}
\usepackage{xcolor}
\usepackage[caption = false]{subfig}
\usepackage{graphicx}
\usepackage{dcolumn}
\usepackage{bm}
\usepackage{hyperref}
\usepackage{xcolor}
\usepackage{amsmath,amsfonts,amssymb}
\usepackage[utf8]{inputenc} 
\usepackage[T1]{fontenc}
\def\apj{ApJ}

\def\mnras{MNRAS}

\def\prd{PRD}

\newcommand{\rmd}{{\rm d}}

\begin{document}
\preprint{FERMILAB-PUB-24-0004-T}
\title{Detecting Dark Matter Substructures on Small Scales with Fast Radio Bursts}
\author{Huangyu Xiao$^{1,2}$}
\email{huangyu@fnal.gov}
\author{Liang Dai$^3$}
\email{liangdai@berkeley.edu}
\author{Matthew McQuinn$^4$}
\email{mcquinn@uw.edu}

\affiliation{$^1$ Astrophysics Theory Department, Theory Division, Fermilab, Batavia, IL 60510, USA}
\affiliation{$^2$Kavli Institute for Cosmological Physics, University of Chicago, Chicago, IL 60637}
\affiliation{$^3$Department of Physics, 366 Physics North MC 7300, University of California, Berkeley, CA
94720, USA}
\affiliation{$^4$Department of Astronomy, University of Washington,  Seattle, WA 98195,USA}

\begin{abstract}
The matter power spectrum is only weakly constrained on subgalactic scales, while physics beyond the Standard Model can leave unique imprints especially on sub-parsec scales. We propose measuring the arrival time difference of Fast Radio Bursts (FRBs) along two adjacent sightlines as a new probe to dark matter substructures on scales down to $\sim 1\,$AU. We discuss two observational scenarios in which it may be possible to place interesting constraints on such models through monitoring repeating FRB sources: 1) By sending radio receivers to space to form a baseline of tens of AU or more and measuring the temporal variation of the arrival time difference between receivers. 2) By measuring the temporal variation of the arrival time difference between two lensed images of one strongly lensed repeater. In both scenarios, obtaining interesting constraints requires correlating the voltage time series to measure the radio-signal arrival time to sub-nanosecond precision. We find that two radio dishes separated by $20\,$AU may be sensitive to the enhancement of small-scale structures at $\sim 10^{-8}M_\odot$ masses in the QCD axion dark matter scenario or from an early epoch of matter-domination with a reheating temperature up to 60 MeV. Other dark matter models such as those composed of $\sim 10^{-13}M_{\odot}$ primordial black holes produced during inflation would also be probed by this method. We further show that a strong lensing situation of multiple images provides an equivalent $\sim 2000\,{\rm AU}\,(\sigma_v/10^3\,{\rm km}{\rm s}^{-1})\,(\delta t/10\,{\rm yr})$ baseline, for a typical velocity of DM substructures $\sigma_v$ and an observational time span $\delta$. This is much more sensitive but with the uncertainty that intervening ISM decoherence may degrade the timing precision and that spatial variation in the FRB emission spot may result in confounding signals. We show that the lensing magnifications of Type Ia supernovea constrain a similar quantity to such FRB timing, with present limits being equivalent to ruling out the same parameter space that would be probed by a $0.14~$AU baseline.

\end{abstract}
\maketitle


\section{Introduction}

Dark Matter (DM) structures on comoving length scales shorter than tens of kiloparsecs all the way down to the solar-system scale have only been weakly constrained by observations
\cite{Viel:2013fqw,Jeong:2014gna,Palanque-Delabrouille:2015pga,Schultz:2014eia,Menci:2016eui,Bovy:2016irg}. 
Enhanced formation of self-gravitating structures on these tiny scales than in the Cold Dark Matter (CDM) scenario may arise from special physical processes that take place in the early Universe. Examples include an early epoch of matter domination \cite{Erickcek:2011us,Fan:2014zua}, an axion arising from a Peccei-Quinn phase transition after inflation \cite{Kolb:1993zz}, a primordial kination era \cite{Redmond:2018xty}, inflationary production of vector DM \cite{Graham:2015rva} and density perturbations from magnetic fields \cite{Adi:2023doe,Ralegankar:2023pyx}.
Therefore, probing such minuscule DM structures would open up a window into the Universe before Big Bang Nucleosynthesis and provide opportunities to discover particle physics beyond the Standard Model.  The only surefire way to detect such structures is via their gravitational effects.

However, those minuscule DM structures, which collapse on comoving wavenumber $k\gtrsim 10$--$10^{10}\,{\rm Mpc}^{-1}$, are notoriously difficult to detect gravitationally.
Femto-, pico-, and microlensing surveys probe masses of  $\sim 10^{-12}M_{\odot}$, but are only sensitive to mass clumps of extremely high densities  ($\gtrsim 10^{15} M_{\odot}/\rm pc^3$) \cite{Kolb:1995bu, Fairbairn:2017sil,Katz:2018zrn}. 
DM substructures in plausible scenarios, such as the so-called ``axion miniclusters'' expected in the QCD axion DM scenario with a post-inflationary Peccei-Quinn phase transition, are orders of magnitude more diffuse \cite{Dai:2019lud, Xiao:2021nkb, Eggemeier:2019khm}. Recently, there have been two proposals for detecting the gravitational effects from DM substructures of asteroid to planet-scale masses that may be sufficiently sensitive to motivated models such as QCD axion miniclusters. One method exploits the imprint of Shapiro time delays and gravitational accelerations imparted by DM substructures in pulsar timing residuals  \cite{Dror:2019twh, Ramani:2020hdo,Lee:2020wfn}. Another proposal suggests the use of perturbing effects collectively caused by many intervening substructures in the microlensing lightcurves of highly magnified extragalactic stars \cite{Dai:2019lud}. Several other methods are sensitive to larger but sub-galactic substructure masses. Astrometric weak lensing signatures imprinted in the proper motion of Galactic stars probe sub-galactic DM subhalos in the Milky Way halo \cite{VanTilburg:2018ykj, Mondino:2020rkn, Mishra-Sharma:2020ynk, Mondino:2023pnc}. Other probes that are sensitive to sub-galactic, super-solar mass DM halos exploit strong lensing of background galaxies \cite{Birrer:2017rpp, Hezaveh:2016ltk, Dai2018Abell370, Dai2020SGASJ1226, Williams2023arXiv230406064W} or gravitational waves \cite{Dai:2018enj, Oguri2022GWLensingWaveEffects, Tambalo2022arXiv221211960T, Lin2023arXiv230404800L}. Additionally, spectral distortions in the cosmic microwave background (CMB) constrain small-scale density fluctuations up to wavenumbers $k=10^4\,\rm Mpc^{-1}$ \cite{Chluba:2012we}. 
As those methods have not yet been used to place significant constraints, it is worthwhile to search for other potential probes.  

In this work, we propose the precision comparison of the arrival times of repeated Fast Radio Burst (FRB) signals along multiple sightlines from the same FRB source as a new direct probe for minuscule DM substructures.  FRBs are radio transients with typically millisecond duration powered by mostly extragalactic sources.
While the origin of FRBs is not yet thoroughly understood, thousands of them have been discovered \cite{CHIMEFRB:2021srp}, with over fifty currently known to repeat \citep[][often on tens of hour timescales]{2023ApJ...947...83C}, and their propagation effects provide a powerful tool to study cosmological physics such as the missing baryon problem, the circumgalactic media of galaxies, the cosmic reionization history \cite{McQuinn:2013tmc,Ravi:2018ose,2019Sci...366..231P,Beniamini:2020ane,Heimersheim:2021meu}, and, in this work, the dark matter substructures. This proposal uses two key aspects of FRBs: they originate from extremely compact sources, and relative timing of the radio signals can be performed to sub-nanosecond precision through coherent analysis of voltage time series \cite{Cassanelli:2021oaw}.

The required multiple sightlines from a single FRB may either connect to a constellation of radio receivers in space separated by astronomical unit (AU)-scales, or arise in a strong lensing situation with multiple lensed images detected by a single receiver. We show that if one source repeatedly emits FRBs, one can measure the arrival time difference between different sightlines and monitor how this varies with time as foreground dark matter substructures move across the field changing the gravitational time delays (and with enough detectors, the gravitational time delays can be separated for a single FRB). This variation can be distinguished from smooth temporal trends caused by mundane kinematic effects, long-term dynamics of galactic structures, or the expansion of the Universe itself \cite{Li:2017mek,Wucknitz:2020spz,Pearson:2020wxb,Tsai:2023tyw}.  We contrast our proposed method with the aforementioned other proposals to detect structures and show that it has the potential to be more sensitive.

Specifically, for radio receivers in space we find that a $0.1\,$AU separation would constrain new parameter space for small-scale dark matter clustering, and that a $20\,$AU separation is potentially sensitive to QCD axion miniclusters and dark matter minihalos produced from an early matter dominated era with a reheating temperature of $60\,$MeV.  
There is already a proposal to do solar system-scale interferometry on FRBs to measure cosmological distances at sub-percent precision \cite{Boone:2022pdz}; our work adds to the science case for this potential experiment. The large fluxes of FRBs allows detections with meter-scale radio dishes that were commonly used on past outer Solar System probes (especially when correlating with a large terrestrial facility to enhance the signal), and the millisecond duration of FRBs makes data transmission back to Earth practical \cite{Boone:2022pdz}. We further show that strong lensing (if the FRB signal remains phase coherent between lensed images) would be a way for nature to generate equivalent baselines of $\sim 2000\,$AU, a concept previously used by several other studies for different applications \cite{Munoz:2016tmg, Dai2017lensedFRBtiming, Lewis:2020asm,Wucknitz:2020spz,Li:2017mek,Xiao:2022hkl,Gao:2023xbi}. 

This work is organized as follows. In Sec.~\ref{sec:2-Dish}, we study the arrival time difference from two separate sightlines under the gravitational influence of intervening DM substructures. In Sec.~\ref{sec:2_image}, we calculate the effect of DM substructures in the scenario of a lensed FRB source with multiple images. In Sec.~\ref{sec:application}, we apply the results in previous sections to physical models such as QCD axion miniclusters. In Sec.~\ref{sec:conclusion}, we summarize our results and present conclusions.


\section{Weak Lensing Effect by Dark Matter Substructures for Two-Dish Configuration}
\label{sec:2-Dish}



\begin{figure}[h!]
    \centering
\includegraphics[width=17cm]{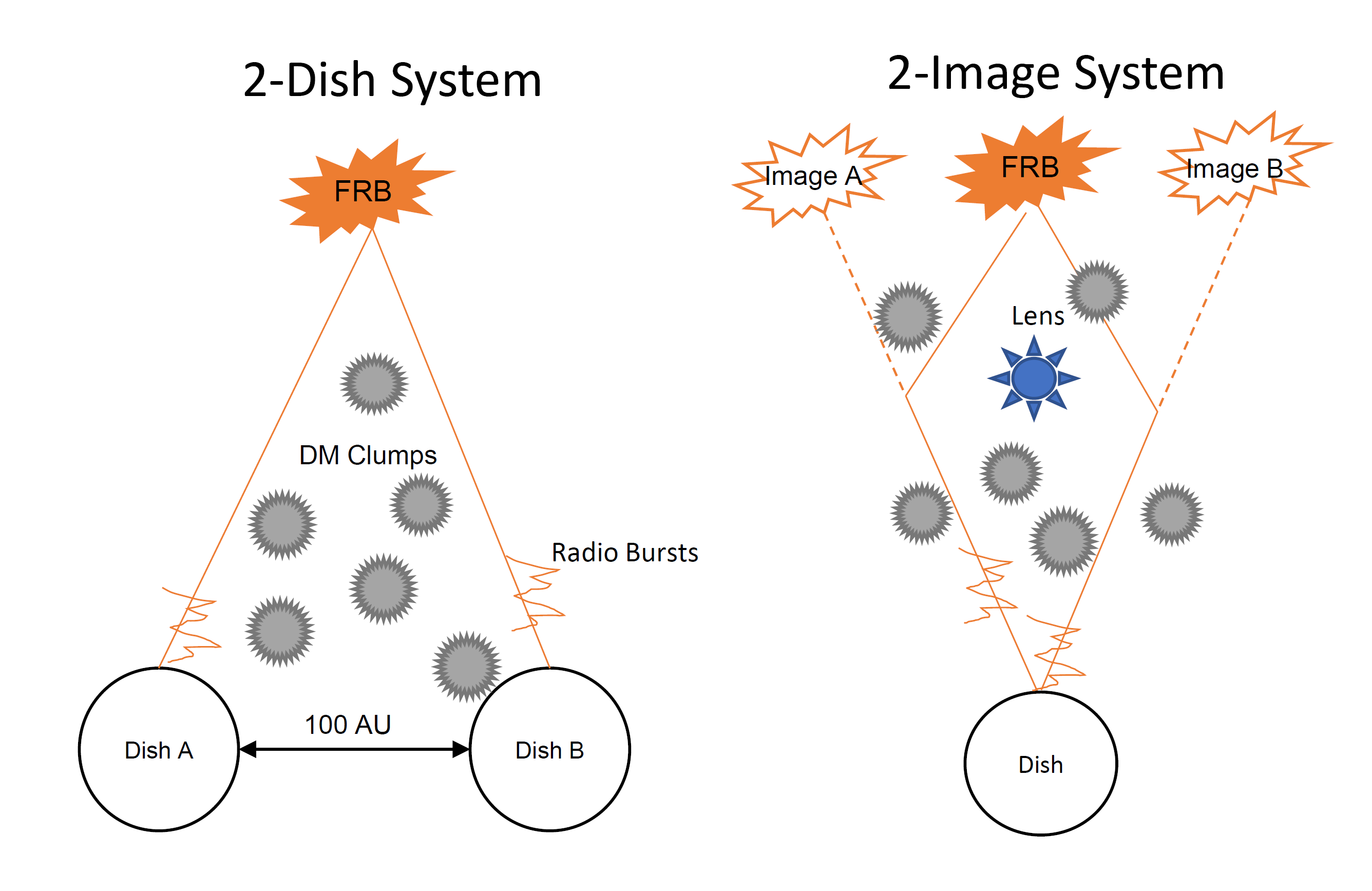}
\caption{
A cartoon picture of the physical scenario we study in this work. On the left side, we anticipate two radio dishes in the solar system separated by 100 AU such that they can observe the same FRB source from two sightlines, which we call a 2-Dish system in this work. Each sightline will experience different Shapiro time-delays from intervening dark matter clumps. Therefore, the FRB electric-field time-series (denoted by the orange pulse curve) will have different time-of-arrivals that can be timed to much better than the frequency of the radio waves.  On the right side, the FRB is strongly lensed, creating two images and we call it a 2-Image system. The two images will experience different Shapiro delays from intervening dark matter substructures and these differences will vary with time. 
}
\label{fig:cartoon}
\end{figure}

Before calculating the effect of DM structures on the arrival time difference along different sightlines, we start with a back-of-the-envelope estimate. As illustrated in the left cartoon of Fig.~\ref{fig:cartoon}, two separate sightlines will experience different time delays induced by dark matter substructures due to different impact parameters.
Consider two radio telescopes separated by 100 AU targeting an extragalactic point source. The number of DM structures within the thin cone formed by the source and the baseline connecting the two dishes is
\begin{equation}
    N_{h}= \frac{\pi \,x_0^2\,D \,\bar{\rho}_m}{3\,M_h}=10 \left(\frac{10^{-6}M_{\odot}}{M_{h}}\right)\left(\frac{D}{1\rm\, Gpc}\right)\left(\frac{x_0}{100\,\rm AU}\right)^2,
\end{equation}
where we assume that all DM substructures have a mass $M_h$, $x_0$ is the separation of the two dishes, $\bar{\rho}_m$ is the mean cosmic DM density today and $D$ is the comoving light-travel distance to the source. In this crude estimate, we also assume that DM substructures are smaller in size than the two-dish separation. As an example, this limit is valid for $x_0 \gtrsim 100\,$AU for the QCD axion miniclusters that collapse after the epoch of matter-radiation equality in the scenario of a post-inflationary Peccei-Quinn phase transition~\cite{Kolb:1993hw}. The fiducial mass scale $M_h \sim 10^{-6}M_\odot$ is motivated by the mass that these miniclusters grow to by $z=0$ in N-body simulations \citep{Xiao:2021nkb}. 
The Shapiro time delay induced by individual dark matter objects is 
\begin{equation}
    \delta t = -GM_h \,{\rm ln}\,(d_{\rm h}/d_{\rm h'}),
\end{equation}
where $d_{\rm h}\sim d_{\rm h^\prime}\sim x_0$ is the typical perpendicular distance from the DM structure to the sightline. Note that $d_{\rm h}$ or $d_{\rm h^\prime}$ is only an estimate of the impact parameter, which has a random value dependent on the exact spatial distribution of DM substructures. Therefore, multiple DM structures cause a variance in the time delay difference between the two sightlines
\begin{equation}\label{eq:t_est}
    \Delta t\sim \sqrt{N_h}\, \delta t = 0.1\,{\rm ns}\,\left(\frac{x_0}{100\,\rm AU}\right)\,\left(\frac{M_{h}}{10^{-6}M_{\odot}}\right)^{1/2} \left( \frac{{\rm ln}\,(d_{\rm h}/d_h^\prime)}{2}\right).
\end{equation}
This simple estimate assumes the DM structure is smaller than $x_0$. The calculations we present later will properly account for the size of DM structures. 
Our method can be compared to two other Galactic methods to measure small DM structures -- pulsar timing arrays (PTAs) and astrometry in the proper motion of Galactic stars. One advantage of our method is the involvement of cosmological path lengths which leads to overall delays that are $\sim $ 100 times larger than in the Galaxy. PTAs are sensitive to residuals of $\sim 100\,$ns \cite{Manchester_2013}, and again we expect the Galactic contribution to our $\Delta t$ to be $\sim 1\%$ of the total delay so our hypothetical experiment should be more sensitive. 
For PTAs, astrophysical systematics such as the spin noise of pulsars and the time variation of electron density (or dispersion measure) induced by the solar wind will have a great impact on the timing precision \cite{Lentati:2016ygu,Tiburzi:2020wuh}. However, FRB timing proposed in this work is intrinsically different from pulsar timing because pulsar timing relies on pulsars being a reference clock while FRB timing uses the difference in arrival times between detectors at the end of different sightlines. By correlating the electric field at two different receivers, the dispersion measure difference, which only enters the correlation as a phase, can be fitted away with a broad frequency band. Therefore, those astrophysical systematics that arise from pulsars or solar wind will not be a concern in our proposal of FRB timing.

To compare with astrometry, the deflection angle is $\Delta \theta \sim \delta t/b$ where $b\sim 100$ AU is the typical impact parameter and $\delta t$ is the Shapiro time delay induced by dark matter substructures. Therefore, a micro-arcsecond precision on the deflection is equivalent to a timing precision of $100$ ns. In addition, astrometry uses dark matter substructures in the galaxy so the number of objects will be smaller than the estimate for cosmological sources.
Future surveys like the Square Kilometre Array (SKA) can reach an angular resolution of 10 $\mu$arcseconds or even $\mu$arcsecond to the brightest stars with 10 years of operation \cite{Fomalont:2004hr}. 
Another advantage of probing minuscule DM structures with cosmological sources such as FRBs is that those DM structures are less disrupted in the low-density intracluster~\cite{Dai:2019lud} or intergalactic space, but more so in the Galaxy due to tidal stripping and stellar disruptions \cite{Kavanagh:2020gcy, Shen:2022ltx}. The microlensing of magnified stars proposal also uses extragalactic sources \cite{Dai:2019lud} but it is challenging to access its sensitivity and currently there is no estimation.

This estimate suggests that a timing precision of $\sim 0.1\,$ns could be interesting for constraining minuscule DM structures. A radio burst signal can be timed to a precision $\sigma_t \sim (2\pi\,\nu)^{-1}\,{\rm SNR}^{-1}$ by cross-correlating the voltage time series recorded at two receivers as long as differential plasma effects do not decorrelate the voltage time series between the two sightlines \cite{Cassanelli:2021oaw}.  Through such coherent techniques, subnanosecond timing is regularly done at gigahertz frequencies. FRBs are bright compact radio sources that are ideal for this timing experiment. If an order-unity fraction of the DM orbits inside tiny structures with masses around $10^{-6}M_{\odot}$, we expect to see random variations in the arrival time difference. The caveat is that the effect can be more suppressed than we have estimated above when the DM structures are larger than the sightline separation. In the following subsections, we will perform detailed calculations for the arrival time difference and present sensitivities on the matter density power spectrum more comprehensively.

\subsection{Arrival time difference introduced by dark matter substructures}

Radio waves passing through the fluctuating gravitational potential $\Phi$ generated by DM substructures are subject to the Shapiro time delay $\Delta t=(2/c^3)\,\int \Phi \,\rmd x$, where $x$ is the comoving coordinate along the line of sight. 
This time delay depends on the realization of random DM substructures. Along two slightly different sightlines the time delay is expected to be also slightly different. This time delay difference has a variance
\begin{equation}\label{eq:Var}
\sigma^2=\frac{4}{c^6}\,{\rm Var}\,\left[\int_0^{D_s}\rmd x\; \Phi(x,\boldsymbol{x}_{\perp})-\int_0^{D_s}\rmd x\; \Phi(x,0) \right],
\end{equation}
where $x_{\perp}$ represents the transverse offset between the two sightlines and $D_s$ is the comoving distance of the FRB source. In the case of an FRB source, the distance along the sightline is cosmological and the transverse offset of interest here is on much smaller length scales comparable to the size of DM substructures.
Therefore, fluctuations in the gravitational potential perpendicular to the sightline are the source of the variance of the arrival time difference. Also, we do not have to include diffraction effects because, for our fiducial specifications, the Fresnel scale $\sim \sqrt{D_s\lambda/(2\pi)}\sim$ AU is at least somewhat smaller than the dark matter substructures or characteristic sightline separations $x_\perp$ we study in this work.
The variance in Eq.~\ref{eq:Var} in principle could be measured if we measure the arrival time difference toward many different one-off FRB sources in the sky. A more practical method would be to target a repeating FRB source and measure the arrival time difference at different times corresponding to the occurrence of multiple FRBs. Since DM substructures are expected to have a velocity component transverse to the slightlines, measuring arrival time difference at different times sample different intervening substructure realizations.
Later in this section, we will show that this temporal variation of the arrival time difference between two sightlines is a powerful probe of DM substructures especially on scales much smaller than kiloparsecs. The comoving-space power spectrum of the gravitational potential $\Phi$ is related to the matter overdensity power spectrum through
\begin{equation}
    P_{\Phi}(k)=\frac{(4\pi G\,\bar{\rho}_{\rm m}\,a^2)^2}{k^4}\,P_{\delta}(k),
\end{equation}
where $\bar{\rho}_{\rm m}$ is the cosmic mean matter density at a given cosmic time.
The variance of the arrival time difference can be expressed in terms of the correlation function of the gravitational potential:
\begin{equation}
    \sigma^2= \frac{8}{c^6}\int_0^{D_s} \rmd x_1\int_0^{D_s}\rmd x_2\; \left( \langle \Phi(x_1,0)\, \Phi (x_2,0)\rangle - \langle \Phi(x_1,0) \,\Phi (x_2,\boldsymbol{x}_{\perp})\rangle \right),
\end{equation}
where $x_1$ and $x_2$ are line-of-sight comoving coordinates. Since correlation functions are invariant under spatial translations, we have used $\langle \Phi(x_1,0)\,\Phi (x_2,0)\rangle=\langle \Phi(x_1,\boldsymbol{x}_{\perp})\,\Phi (x_2,\boldsymbol{x}_{\perp})\rangle$ to simplify the expressions. The correlation function is the Fourier transform of the power spectrum
\begin{equation}
\langle \Phi(x_1,0) \,\Phi (x_2,\boldsymbol{x}_{\perp})\rangle=\int \frac{\rmd^3k}{(2\pi)^3}\,P_{\Phi}(k)\,{\rm exp}\left[-i\,k_{\parallel}\,(x_1-x_2)-
i\,\boldsymbol{k}_{\perp}\cdot \boldsymbol{x}_{\perp} \right],
\end{equation}
where we have decomposed the wave vector $\boldsymbol{k}$ into the line-of-sight component $k_\parallel$ and the transverse component $\boldsymbol{k}_\perp$. The tremendous length scale along the line of sight means that large $k_{\parallel}$ values contribute little to the integral. Thus, under the so-called Limber's approximation, we may neglect the $k_{||}$ dependence of $P_\Phi$ and evaluate the line-of-sight integral as a Dirac $\delta$-function  \cite{1953ApJ...117..134L}:
\begin{equation}
    \langle\Phi(x_1,0) \,\Phi (x_2,\boldsymbol{x}_{\perp})\rangle =  \int \frac{\rmd^2\boldsymbol{k}_{\perp}}{(2\pi)^2}\,P_{\Phi}(k_{\perp})\,\delta_D(x_1-x_2)\,e^{-i \boldsymbol{k}_{\perp}\cdot \boldsymbol{x}_{\perp}}.
\end{equation}
Using this result, the time delay difference has a variance
\begin{equation}\label{eq:sigma_v1}
\begin{split}
\sigma^2 & = \frac{8}{c^6}\,(4\pi G\,\bar{\rho}_{\rm m})^2\int^{z_s}_0\frac{c\,\rmd z}{H(z)}\int \frac{\rmd^2\boldsymbol{k}_{\perp}}{(2\pi)^2}\frac{1}{k_{\perp}^4}P_{\delta}(k_{\perp},z)\,\left(1-e^{-i \boldsymbol{k}_{\perp}\cdot \boldsymbol{x}_{\perp}} \right),  \\
     &=\frac{8}{c^6}\,(4\pi G\,\bar{\rho}_{\rm m})^2\int^{z_s}_0\frac{c\,\rmd z}{H(z)}\int \frac{\rmd k_{\perp}}{2\pi}\frac{1}{k_{\perp}^3}P_{\delta}(k_{\perp},z)\,\left(1-J_0(k_{\perp}x_{\perp})\right),
\end{split}
\end{equation}
for an FRB source at redshift $z_s$. The second line is derived by integrating out the angular component of $\boldsymbol{k}_{\perp}$, which gives the Bessel function. 

If the source of the fluctuating gravitational potential has a relatively long spatial correlation length such that $k_{\perp}\,x_{\perp}\ll 1$, we may Taylor-expand the Bessel function as $1-J_0(k_{\perp}x_{\perp}) = (k_{\perp}x_{\perp}) ^2/4-(k_{\perp}x_{\perp})^4/64 + \cdots$.
Indeed, the leading term in this expansion which scales as $\propto x_\perp^2$ is degenerate with the angular position of the source in the absence of a fluctuating gravitational potential. 
This corresponds to an arrival time difference that scales linearly with $x_\perp$, or equivalently one that varies linearly with the angle. The contribution to the variance comes from a linear delay across the field, and a linear delay is indistinguishable from a slightly different direction of the source. We cannot break the degeneracy without the measurement of FRB repeaters that contain time variation signals, which forces us to consider higher order terms on $x_\perp$. 

We therefore have to study the next-order term which scales as $\propto x_\perp^4$.
In fact, in many cosmology scenario of interest potential fluctuations of low wavenumbers make important contributions to Eq.~\ref{eq:sigma_v1}, so that Taylor-expansion of the Bessel function is justified. The consideration of the next-order term leads to the following quartic contribution:\footnote{This quartic contribution from small structures is degenerate with the effect of curvature of the wavefront \cite{Boone:2022pdz}.  However, wavefront curvature can be largely removed using our precise cosmological constraints to predict distance from the FRB redshift or by concentrating on more distant FRBs.  Furthermore, our effect creates a quadrupolar delay in angle with respect to the baseline vector that can be distinguished from wavefront curvature with enough detectors.} 
\begin{equation}\label{eq:sigma_constant}
    \sigma_4^2=\frac{1}{8\,c^6}\,\left(4\pi G\,\bar{\rho}_{\rm m} \right)^2\,\int^{z_s}_0\frac{c\,\rmd z}{H(z)}\int\, \frac{\rmd k_{\perp}}{2\pi}\,k_{\perp}\,P_{\delta}(k_{\perp},z)\,x_{\perp}^4.
\end{equation}

For a crude estimate in the case of an FRB source, we may assume that the matter power spectrum does not vary substantially with redshift and the FRB is at a comoving distance $D_{\rm s}$. 
The transverse offset between the two sightlines $x_{\perp}$ has a maximal value $x_0$ on Earth and vanishes on the FRB source; it varies along the sightline according to $x_\perp = x_0\,D(z)/D(z_s)$, where $D(z)$ is the comoving distance from earth at redshift $z$. With these simplifications, we derive the expression
\begin{equation}\label{eq:sigma_4}
    \sigma_4^2\approx\frac{1}{40\,c^6}(4\pi G\,\bar{\rho}_{\rm m})^2\,x_0^4\,D_{\rm s}\,\int \frac{\rmd k_{\perp}}{2\pi}\,k_{\perp}\,P_{\delta}(k_{\perp},0).
\end{equation}
A larger transverse separation $x_0$ will lead to a more detectable effect (the time delay signal is proportional to $x_0^2$).
The overall contribution of this higher-order term (the quadrupole term or the circularly symmetric term) in Eq.~\ref{eq:sigma_4} where the delay scales quadraticly in $|x_\perp|$ is suppressed due to the small dish separation.
However, this suppression can be alleviated by looking at the temporal change on the arrival time difference if we detect repeating FRBs, as we will see in the next subsection. In other words, we can use the measurement of repeating FRBs to break the degeneracy on the leading order with source angle.\footnote{This does additionally require an absolute measurement of the angle to the FRB, which necessitates knowing the overall rotational orientation of the array and not just the relative positions of the receivers. This requires using other FRBs to set the reference frame or, possibly, precise two-way ranging to measure the distances between the array elements and terrestrial receivers whose position is extremely well modeled.} Additionally, the movement of dark matter substructures can effectively increase the baseline length and the variance of the arrival time difference can be enhanced by a factor of  $(x_0/[ v\, \delta t])^2$, where $\delta t$ is the duration over which the repeating FRB is observed and $v$ is the velocity of dark matter substructures. Even though observing the quartic contribution in Eq.~\ref{eq:sigma_constant} is less sensitive except for extremely large baselines, it has the advantage that it does not require the FRB to repeat and can be used as a cross check of the method we discuss next. 


\subsection{Time variance of weak lensing effect with repeating FRBs}


 We now extend our calculation to time variable signals.  We can measure the arrival time differences between the two sightlines at multiple epochs by targeting a repeating FRB source.
This will be the primary observable we study here.
Currently fifty FRB sources have been found to repeat on $\mathcal{O}(10\,{\rm hr})$ timescales, and the detected number of repeating sources is growing quickly \cite{CHIMEFRB:2023myn}.
The change in the gravitational potential with time arises from the inevitable motion of the DM substructures; over a time intervel $\delta t$, the effect of this motion is equivalent to shifting the transverse spatial coordinate of $\Phi$ the sightline probes by an amount $\boldsymbol{v}_{\perp}\,\delta t$. 
We may assume the dominant velocity of the DM substructures relative to the sightlines results from the local motion of the large-scale structure, which is coherent over a comoving distance $\sim$ 50 Mpc.
We can calculate the variance of the arrival time difference over a time interval $\delta t$: 
\begin{equation}
\sigma_t^2=\frac{4}{c^6}{\rm Var}\, \left[\int_0^{D_s} \rmd x\, \Phi(x,\boldsymbol{x}_{\perp})-\int_0^{D_s}\rmd x \,\Phi(x,\boldsymbol{0})-\int_0^{D_s} \rmd x\, \Phi(x,\boldsymbol{x}_{\perp}+\boldsymbol{v}_{\perp}\delta t)+\int_0^{D_s} \rmd x \,\Phi(x,\boldsymbol{v}_{\perp}\delta t) \right],
\end{equation}
The above expression vanishes if $\delta t=0$.
For a unique transverse velocity $\boldsymbol{v}_{\perp}$, we can calculate
\begin{equation}
    \begin{split}
        \sigma_t^2(v_{\perp})= & -\frac{4}{c^6}\int_0^{D_s} \rmd x_1\int_0^{D_s}\rmd x_2\,\Bigl [ 4\,\langle \Phi(x_1,\boldsymbol{0})\, \Phi (x_2,\boldsymbol{x}_{\perp})\rangle+4\,\langle\Phi(x_1,\boldsymbol{0})\, \Phi (x_2,\boldsymbol{v}_{\perp}\delta t)\rangle-4\,\langle \Phi(x_1,\boldsymbol{0}) \,\Phi (x_2,\boldsymbol{0})\rangle \\
        &-2\,\langle \Phi(x_1,\boldsymbol{0})\, \Phi (x_2,\boldsymbol{x}_{\perp}+\boldsymbol{v}_{\perp}\delta t)\rangle-2\,\langle \Phi(x_1,0) \,\Phi (x_2,-\boldsymbol{x}_{\perp}+\boldsymbol{v}_{\perp}\delta t)\rangle
        \Bigr ]\\
        =& \frac{4}{c^6}(4\pi G\bar{\rho}_{\rm m})^2\int_0^{z_s}\frac{c\,\rmd z}{H(z)}\int \frac{\rmd^2\boldsymbol{k}_{\perp}}{(2\pi)^2}\frac{1}{k_{\perp}^4}P_{\delta}(k_{\perp},z)\,\Bigl [ 4 + 2\,e^{-i \boldsymbol{k}_{\perp}\cdot (\boldsymbol{x}_{\perp}+\boldsymbol{v}_{\perp}\delta t)}+2\,e^{-i \boldsymbol{k}_{\perp}\cdot (-\boldsymbol{x}_{\perp}+\boldsymbol{v}_{\perp}\delta t)}\\
        & -4\,e^{-i \boldsymbol{k}_{\perp}\cdot \boldsymbol{x}_{\perp}} - 4\,e^{-i \boldsymbol{k}_{\perp}\cdot \boldsymbol{v}_{\perp}\delta t} \Bigr ] \\
        = & \frac{8}{c^6}(4\pi G\bar{\rho}_{\rm m})^2\int_0^{z_s}\frac{c\,\rmd z}{H(z)}\int \frac{\rmd^2\boldsymbol{k}_{\perp}}{(2\pi)^2}\frac{1}{k_{\perp}^4}\,P_{\delta}(k_{\perp},z)\,\left[ \left(1-e^{i\boldsymbol{k}_{\perp}\cdot \boldsymbol{x}_{\perp}}\right)\,\left(1-e^{-i\boldsymbol{k}_{\perp}\cdot \boldsymbol{v}_{\perp}\delta t} \right) + \left(1-e^{i\boldsymbol{k}_{\perp}\cdot \boldsymbol{x}_{\perp}}\right)\,\left(1-e^{i\boldsymbol{k}_{\perp}\cdot \boldsymbol{v}_{\perp}\delta t} \right) \right].
    \end{split}
\end{equation}
In reality, the DM substructure velocity $\boldsymbol{v}_\perp$ involved in the above expression varies along the sightline. The peculiar motion of the large-scale structure is coherent over distances of tens of megaparsecs, while the FRB source is at a cosmological distance. The sightline therefore samples different coherent patches of the large-scale velocity field. This justifies taking the statistical average of the above expression with respect to the distribution of $\boldsymbol{v}_\perp$. Accounting for large-scale linear matter overdensity modes, the velocity is drawn from a Gaussian distribution with a standard deviation of $\sigma_v \approx 600 \sqrt{2/3}$ km~s$^{-1}$ at low redshifts, where $\sqrt{2/3}$ accounts for only the transverse component of the velocity vector~\footnote{We neglect the contribution from local collapsed structures here, which would increase somewhat the dispersion.}.
By integrating $\sigma_t^2(v_{\perp})$ over the Gaussian distribution of $v_{\perp}$, we derive:
\begin{equation}\label{eq:sigmat}
\begin{split}
          \sigma_t^2 &= \frac{16}{c^6}(4\pi G\bar{\rho}_{\rm m})^2\int_0^{z_s}\frac{c\,\rmd z}{H(z)}\int \frac{\rmd^2 \boldsymbol{k}_{\perp}}{(2\pi)^2}\frac{1}{k_{\perp}^4}P_{\delta}(k_{\perp},z) \left(1-e^{i\boldsymbol{k}_{\perp}\cdot \boldsymbol{x}_{\perp}} \right)\,\int \frac{\rmd^2\boldsymbol{v}_{\perp}}{2\pi \sigma_v^2}\,\exp\left(-\frac{v_{\perp}^2}{2\sigma_v^2}\right)\left(1-e^{-i\boldsymbol{k}_{\perp}\cdot \boldsymbol{v}_{\perp}\delta t}\right)\\
          &=\frac{16}{c^6}(4\pi G\bar{\rho}_{\rm m})^2\int_0^{z_s}\frac{c\,\rmd z}{H(z)}\int \frac{\rmd^2\boldsymbol{k}_{\perp}}{(2\pi)^2}\frac{1}{k_{\perp}^4}P_{\delta}(k_{\perp},z) \left( 1-e^{i\boldsymbol{k}_{\perp}\cdot \boldsymbol{x}_{\perp}} \right)\,\left(1-e^{-{k}^2_{\perp}\sigma_v^2\delta t^2/2}\right)\\
          &=\frac{16}{c^6}(4\pi G\bar{\rho}_{\rm m})^2\int_0^{z_s}\frac{c\,\rmd z}{H(z)}\int \frac{\rmd k_{\perp}}{2\pi}\frac{1}{k_{\perp}^3}P_{\delta}(k_{\perp},z) (1-J_0(k_{\perp} x_{\perp}))\,\left( 1-e^{-k^2_{\perp}\sigma_v^2\delta t^2/2}\right),
\end{split}
\end{equation}
In the limit of $k_{\perp} x_{\perp}\ll 1$,
which applies to scenarios in which small-scale DM substructures have sizes much larger than the transverse separation between the two sightlines, the variance has a simpler expression
\begin{equation}
    \sigma_t^2=\frac{4}{c^6}(4\pi G\bar{\rho}_{\rm m})^2\int_0^{z_s}\frac{c\,\rmd z}{H(z)}\int \frac{\rmd k_{\perp}}{2\pi}\frac{1}{k_{\perp}}P_{\delta}(k_{\perp},z)\,\left(1-e^{-k^2_{\perp}\sigma_v^2\,\delta t^2/2} \right)\,x_{\perp}^2.
    \label{eqn:sigmat}
\end{equation}
If potential fluctuation modes of short wavelengths are not a dominant contribution, i.e. if $k_{\perp}\sigma_v\,\delta t\ll 1$, the time delay difference $\Delta t$ between the two sightlines as a function of time $t$ can be expanded as a Taylor series $\Delta t= a_0+a_1\,t+a_2\,t^2+...$. The expansion coefficients, encoding information about the DM substructures, can be measured accurately with sufficiently many well-measured FRB repetitions. The leading term (constant term $a_0$), independent of $\delta t$, is described by Eq.~\ref{eq:sigma_v1} when expanding the Bessel function, which is degenerate with the source angular location. The next term is related to the variance of the time derivative on the arrival time difference, which is expressed as
\begin{equation}\label{eq:sigmat'}
\begin{split}
        \sigma_{ t^{\prime}}^2&=\frac{4}{c^6}\int\frac{\rmd^2\boldsymbol{v}_{\perp}}{2\pi \sigma_v^2}\,\exp\left(-\frac{\boldsymbol{v}_{\perp}^2}{2\sigma_v^2}\right){\rm Var}\left[\frac{\partial}{\partial t}\left(\int_0^{D_s} \rmd x \,\Phi(x,x_{\perp}+v_\perp t)-\int_0^{D_s}\rmd x\,\Phi(x,v_\perp t)\right)\right]\\
        &=\frac{8}{c^6}(4\pi G\bar{\rho}_{\rm m})^2\int_0^{z_s}\frac{c\,\rmd z}{H(z)}\int \frac{\rmd k_{\perp}}{2\pi}\frac{1}{k_{\perp}^3}P_{\delta}(k_{\perp},z)(1-J_0(k_{\perp}x_{\perp}))\int\frac{\rmd^2\boldsymbol{v}_{\perp}}{2\pi \sigma^2}\,\exp\left(-\frac{\boldsymbol{v}_{\perp}^2}{2\sigma_v^2}\right)\,(\boldsymbol{k}_{\perp}\cdot \boldsymbol{v}_{\perp})^2 \\
        &=\frac{8}{c^6}(4\pi G\bar{\rho}_{\rm m})^2\sigma_v^2\int_0^{z_s}\frac{c\,\rmd z}{H(z)}\int \frac{\rmd k_{\perp}}{2\pi}\frac{1}{k_{\perp}}P_{\delta}(k_{\perp},z)(1-J_0(k_{\perp}x_{\perp})).\\
\end{split}
\end{equation}
The integrand here is as sensitive to higher wavenumber (larger $k_\perp$) modes of potential fluctuations as the higher order term (the term that is $\propto x_\perp^4$ when expanding the Bessel function $J_0(k_\perp x_\perp$).) in Eq.~\ref{eq:sigma_constant}.
However, it can be shown that the time derivative term will be more sensitive to higher wavenumber modes than the term described by Eq.~\ref{eq:sigma_constant} if $\sigma_v\delta t>x_0$. In other words, if the transverse spatial separation between the two sightlines is limited by the available technologies to realize a long spatial baseline, we can compensate for that by observing over a long time baseline, during which the sightlines sweep DM substructures across a much longer distance. For reference, $\sigma_v\,\delta t\approx 100$ AU if we monitor a repeating FRB source for one year $\delta t =$ 1 yr, which is equivalent to a spatial baseline between the Earth and the outer Solar System. One can compare the time varying signal in Eq.~\ref{eq:sigmat'} to the higher order term in Eq.~\ref{eq:sigma_constant} and find that they reach the same sensitivity to small-scale potential fluctuations when $\sigma_v\delta t = x_0$. Therefore, we can be more sensitive to small-scale DM structures by exploiting time variation of the arrival time difference with 10 years of observation of repeating FRBs, compared to the static method that uses higher order terms since it is very challenging to realize a 1000 AU spatial baseline.
The measurement of the time derivative can be even more accurate if a large number of repeating events are observed for the same FRB source and a common trend is measured. 
The caveat is that many other physical effects, including the expansion of the Universe in real time \cite{Li:2017mek}, might cause the arrival time difference to vary linearly with time. Therefore, arrival time differences that vary over shorter timescales would be more interesting signals of small-scale DM structures. 


While the linear trend induced by DM substructures must dominate other confounding effects to be detectable, assuming this is the case then if the signal falls primarily in only one logarithmic bin of wavenumber $k$, we can measure the matter power spectrum on that scale to a precision
\begin{equation}\label{eq:Pk_t1}
    \delta P_{\delta}(k) \approx \frac{2\pi\,\delta t_{m}^2/N}{\frac{1}{c^6}(4\pi G\bar{\rho}_{\rm m})^2\, x_0^2\, D_{\rm s}\,\sigma_v^2\,\delta t^2\,{k^2}},
\end{equation}
where $N$ is the number of repeating events minus 1 (we need to measure at least 1 burst to fit the mean value of the arrival time difference) and $\delta t_m$ is the FRB timing precision, which we argue can reach subnanosecond at high frequencies by cross-correlating voltage time series. Note that the above expression is valid only if $\sigma_v \delta t \, k \ll 1$. 

Consider fiducial parameter values corresponding to an interferometry setup on the Solar System scale. We find the numerical results
\begin{equation}\label{eq:Pk_t1es}
    \frac{\delta P_{\delta}(k)\,k^3} {2\pi^2} \approx 8.5\times 10^2 \left(\frac{N}{100}\right)^{-1}  \left(\frac{k}{\rm pc^{-1}} \right) \left(\frac{\delta t}{10 \,\rm yr} \right)^{-2} \left(\frac{\delta t_m}{0.1\,\rm ns} \right)^{2}  \left(\frac{D_s}{1\,\rm Gpc} \right)^{-1} \left(\frac{x_0}{100\,\rm AU}\right)^{-2}.
\end{equation}
As we can see in Appendix.~\ref{app:halo_model}, the amplitude of the dimensionless power spectrum of axion miniclusters at present day can easily reach $10^6$ since it is already order unity at matter-radiation equality and will continue to grow according to the calculations using the halo model.
The sensitivity for sufficient high-$k$ Fourier modes with $\sigma_v\,\delta t \,k_{\perp}\gtrsim 1$ cannot be calculated from the previous expressions. For that, we have to study the temporal power spectrum of the arrival time difference in Eq.~\ref{eqn:Ptomega}, which will be discussed later in Section \ref{subsec:temporal}.

In principle, we will not only measure the rate at which the time delay difference varies linearly with time, but also measure the correction piece with a quadratic dependence on time. Such measurement will allow us to focus on contributions from higher $k$ modes of potential fluctuations since higher $k$ modes make more contributions to the integral. The variance of the quadratic derivative can be expressed as:
\begin{equation}\label{eq:sigmat2}
    \begin{split}
       \sigma_{t^{\prime\prime}}^2&=\frac{4}{c^6}\int\frac{\rmd^2\boldsymbol{v}_{\perp}}{2\pi \sigma_v^2}\,{\rm exp}\left(-\frac{v_{\perp}^2}{2\sigma_v^2}\right){\rm Var}\left[ \frac{\partial^2}{\partial t^2}\left(\int_0^{D_s} \rmd x \, \Phi(x,x_{\perp}+v_\perp t)-\int_0^{D_s}\rmd x \,\Phi(x,v_\perp t)\right)\right]\\
        &=\frac{24}{c^6}(4\pi G\bar{\rho}_{\rm m})^2\sigma_v^4\int_0^{z_s}\frac{c\,\rmd z}{H(z)}\int \frac{\rmd k_{\perp}}{2\pi}{k_{\perp}}P_{\delta}(k_{\perp},z)(1-J_0(k_{\perp}x_{\perp})).
    \end{split}
\end{equation}
Again, if we restrict to the regime $k_{\perp}x_{\perp}\ll 1$, the above expression can be approximated
\begin{equation}
    \begin{split}
       \sigma_{t^{\prime\prime}}^2
        &\approx\frac{6}{c^6}(4\pi G\bar{\rho}_{\rm m})^2\sigma_v^4\int_0^{z_s}\frac{c\,\rmd z}{H(z)}\int \frac{\rmd k_{\perp}}{2\pi}{k_{\perp}^3}P_{\delta}(k_{\perp},z)\,x_{\perp}^2.
    \end{split}
\end{equation}
The above result shows that the quadratic time derivative term will be more sensitive to higher $k$ modes. Adopting the same assumption that the $P_\delta$ is only nonzero in a single logarithmic bin of $k$, we derive the following sensitivity to the matter power spectrum from $\sigma_{t^{\prime\prime}}$:
\begin{equation}\label{eq:PK_t2}
    \delta P_\delta(k)\approx \frac{2\pi\, \delta t_{\rm m}^2/N}{\frac{3}{c^6}(4\pi G\bar{\rho}_{\rm m})^2\,x_0^2\,D_{\rm s}\,\sigma_v^4 \,\delta t^4\,{k^4}}.
\end{equation}
Here $N$ refers to the number of repeating events minus 2 (we need to measure 2 bursts to fit both the mean value and the first derivative of the arrival time difference). Again, this result cannot be applied to modes of arbitrarily high $k$ but must be restricted to $k_{\perp}\,\sigma_v\,\delta t\sim 1$. We will study the temporal power spectrum (Eq.~\ref{eqn:Ptomega}) of the arrival time difference to overcome the limitation of this expansion scheme based on the wavenumber. For now, the sensitivity for the matter power spectrum numerically evaluates to:
\begin{equation}\label{eq:Pk_t2es}
    \frac{\delta P_{\delta}(k)\,k^3} {2\pi^2} \approx 1.1\times 10^6\left(\frac{N}{100}\right)^{-1}  \left(\frac{k}{\rm 10 pc^{-1}} \right)^{-1} \left(\frac{\delta t}{10 \,\rm yr} \right)^{-4} \left(\frac{\delta t_m}{0.1 \rm ns} \right)^{2}  \left(\frac{D_s}{1 \rm Gpc} \right)^{-1} \left(\frac{x_0}{100\rm AU}\right)^{-2}.
\end{equation}
Noticeably, the sensitivity based on the term with a quadratic time dependence is better for higher $k$ modes.

\begin{figure}[h!]
    \centering
        \begin{minipage}{0.48\textwidth}
        \centering
        \includegraphics[width=\textwidth]{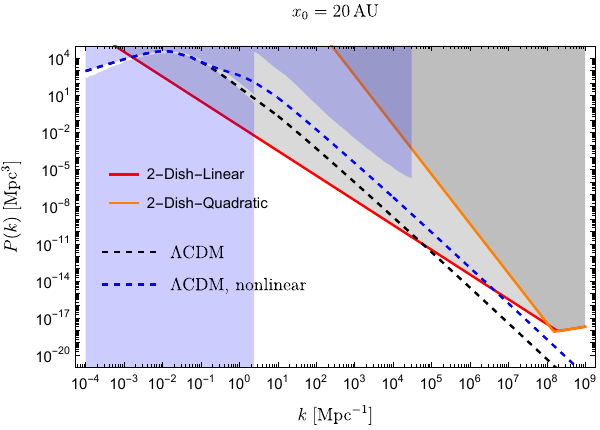}
    \end{minipage}\hfill
    \begin{minipage}{0.48\textwidth}
        \centering
        \includegraphics[width=\textwidth]{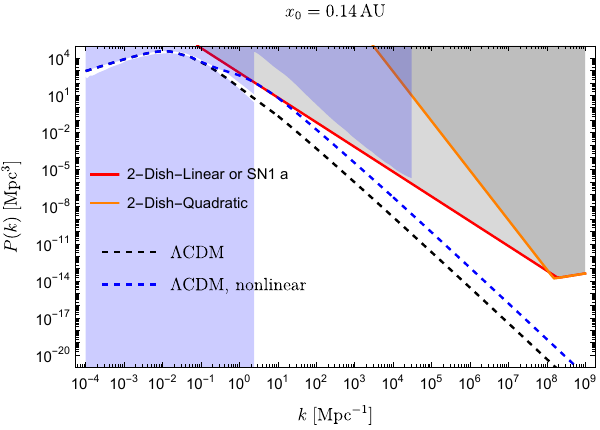}
    \end{minipage}
\caption{The sensitivity of the 2-Dish system to the nonlinear matter power spectrum from measurements of the time-varying arrival time difference of repeating FRBs with antennas separated by $x_0=20$ AU (left panel) or by $x_0=0.14$ AU (right panel). The blue-shaded region has been excluded by current observations such as CMB \cite{Hu:1994bz,Planck:2015sxf}, galaxy formation, and Lyman-alpha forest \cite{2011MNRAS.413.1717B}. The blue triangular exclusion region at higher wavenumber is from constraints on CMB spectral distortions \cite{Hu:1994bz}.
The gray region would potentially be excluded by this hypothetical 2-Dish system. Note that the 2-Dish system probes the nonlinear matter power spectrum in the late Universe while observations like CMB distortions probe the matter power spectrum in the early Universe when it was still linear.
On the right panel, the gray region should already be excluded by supernova lensing since it is equivalent to a 0.14 AU baseline, as discussed in Appendix.~\ref{app:sn_lensing}.
The linear matter power spectrum in $\Lambda$CDM is plotted as the black dashed curve as a reference and the nonlinear power spectrum is the blue dashed curve, which uses the HaloFit model \cite{Takahashi:2012em,Smith:2002dz}. The red solid curve is the detection threshold by measuring the first derivative of time delay difference, whose variance is calculated in Eq.~\ref{eq:Pk_t1} and the orange solid curve represents the sensitivity by measuring the second derivative of time delay difference calculated in Eq.~\ref{eq:PK_t2}. The matter power spectrum above the solid curves is detectable.
The $\Lambda$CDM contribution is still significant in the first derivative measurement but much more suppressed in the second derivative term.
The fiducial values of parameters in this plot are $\delta t_m = 0.1$ ns, $D_{\rm s}= 3$ Gpc, $\sigma_v=500$ km/s, $\delta t=10$ yr and $N=100$.
At high $k$ modes, the slope of the solid curves turns around ($\propto k^{1/2}$), which is caused by the failure of the expansion of time delay difference  $\Delta t= a_0+a_1\,t+a_2\,t^2+...$ when $k\,\sigma_v\delta t\sim 1$, where all order of derivatives will matter and one should study the temporal power spectrum in (c.f. Eq.~\ref{eqn:Ptomega}) to obtain the sensitivity.
}
    \label{fig:Pk_constraint1}
\end{figure}

In Fig.~\ref{fig:Pk_constraint1}, we plot the sensitivity on the matter power spectrum for the 2-Dish configuration, from measuring $\rmd \Delta t/\rmd t$ and $\rmd^2 \Delta t/\rmd t^2$, where $\Delta t$ is the arrival time difference along two sightlines.  The 2-Dish system refers to radio receivers in space that create separate sightlines for the same FRB source and separated by $x_0=20$AU (left) and $x_0 = 0.14$AU (right).
The sensitivity curves are calculated for the following fiducial parameter values: FRB timing precision $\delta t_m = 0.1$ ns, comoving distance to the source of $D_{\rm s}= 3$ Gpc (or $z\approx 1$), characteristic dispersion of peculiar velocity of the DM structures $\sigma_v=500$ km/s, time span of observing repeating FRB sources $\delta t=10$ yr, and the number of $\delta t_m$-timed radio bursts is $N=100$. 
The matter power spectrum expected from the standard $\Lambda$CDM cosmology is also plotted in Fig.~\ref{fig:Pk_constraint1}, though we must emphasize that the high-$k$ regime ($k> 10\,\rm Mpc^{-1}$) is poorly constrained. The black dashed curve in Fig.~\ref{fig:Pk_constraint1} is the linear matter power spectrum in $\Lambda$CDM cosmology, while the blue dashed curve is the nonlinear matter power spectrum obtained from the HaloFit model \cite{Takahashi:2012em,Smith:2002dz}. Note that the nonlinear power spectrum has been extrapolated to an extremely high wavenumber regime, well beyond the wavenumbers the HaloFit model has not been calibrated to simulations.
Our observational probe constrains the nonlinear power spectrum. The blue region is excluded by current observations from CMB \cite{Hu:1994bz,1994ApJ...420..439M,Fixsen:1996nj,Planck:2015sxf}, and Lyman-alpha forest \cite{2011MNRAS.413.1717B} (The blue triangular region is currently excluded by limits on CMB spectral distortions; \citealt{}). The gray region is the sensitivity forecast for the 2-Dish system we study in this work. Technically this system requires four dishes to constrain their positions via GPS techniques, although the other dishes are not necessarily as large as the ``2-Dish'' we are relying on to detect FRBs. The sensitivities of the 2-Dish system to $\rmd \Delta t/\rmd t$ and $\rmd^2 \Delta t/\rmd t^2$ are respectively calculated using Eq.~\ref{eq:Pk_t1} and Eq.~\ref{eq:PK_t2}.

We find that the current limit on supernova lensing magnifications provides a limit on the matter power spectrum that is equivalent to the 2-Dish configuration with a $x_0=0.14$ AU baseline, as discussed in Appendix.~\ref{app:sn_lensing}. Thus, the gray region above the red curve in the right panel in Fig.~\ref{fig:Pk_constraint1} should be considered as a current liimit on the nonlinear matter power spectrum. 
Supernovae lensing already provides comparable if not better sensitivity to the CMB spectral distortion for a wide range of wavenumber. It is worth noting that the comparison to CMB spectral distortions is not so direct as the observable we proposed in this work is at low redshift ($z\sim 1$) while CMB distortions are sensitive to matter power spectrum before recombination, which is still in the linear regime. Therefore our proposal will probe the nonlinear matter power spectrum, which requires some forward modeling to extrapolate the primordial linear matter power spectrum to the current Universe.  Section~\ref{sec:application} discusses this modeling in the context of motivated smalls-scale enhancements in the matter power. 


Large scale structures in the $\Lambda$CDM cosmology that have 10s or even hundreds of wavelengths still affect the first time derivative of arrival time difference from the gradient field of their gravitational potential (as seen by the 2-Dish linear running somewhat parallel to the matter power spectrum in Fig.~\ref{fig:Pk_constraint1}). This means that baselines much longer than  $x_0=0.14\,$AU will detect the $\Lambda$CDM contribution and not be able to probe the potential high-$k$ enhancements we are interested in.  The way to avoid this large-scale sensitivity is to measure the second time derivative of the arrival time difference, as it will only be sensitive to the matter overdensity modes at higher $k$ with the $\Lambda$CDM contribution efficiently filtered out. The sensitivity to the second time derivative is shown by the orange curves in Fig.~\ref{fig:Pk_constraint1}.

The expansion of the time delay difference in terms of the first derivatives, the second derivative, or even higher-order derivatives, is only valid when the dominant modes satisfy $k\,\sigma_v\,\delta t \ll 1$. When $k$ is larger than $1/(\sigma_v\,\delta t)$, the temporal power spectrum of the arrival time difference is the appropriate observable that preserves the full information. This change over in constraining method is reflected by the kinks in the sensitivity curves, where the higher $k$ beyond the kink uses the temporal power spectrum to constrain the matter power spectrum (which will be discussed in the next subsection). 
Practically, the highest wavenumber that such FRB timing may probe is given by repeating rate of FRBs, which for some of the most prolific repeaters is $\sim 1/(10)~$hr$^{-1}$ \citep{CHIMEFRB:2023myn}, divided by the velocity of the dark matter substructure, $\sigma_v$, which yields $k\sim 10^{12} \,\rm Mpc^{-1}$.


\subsection{Temporal Power Spectrum}\label{subsec:temporal}

Our approach of measuring derivatives of the arrival times brings in higher wavenumber information with increasing order.  A different approach that captures all this information in a single statistic is to measure the temporal correlation function or power spectrum of the time delay measurements. The temporal correlation function of the time delay difference is
\begin{equation}
\begin{split}
        \langle\Delta t(0)\,\Delta t(\delta t)\rangle&=\int\frac{\rmd^2\boldsymbol{v}_{\perp}}{2\pi \sigma_v^2}\,{\rm exp}\left(-\frac{v_{\perp}^2}{2\,\sigma_v^2}\right) \times\\
        &
        \qquad\left\langle \left(\int_0^{D_s} \rmd x \,\Phi(x,x_{\perp})-\int_0^{D_s}\rmd x\, \Phi(x,0)\right)\left(\int_0^{D_s} \rmd x \,\Phi(x,x_{\perp}+v_{\perp}\delta t)-\int_0^{D_s}\rmd x \,\Phi(x,v_{\perp}\delta t)\right)\right\rangle\\
         &=\frac{8}{c^6}(4\pi G\bar{\rho}_{\rm m})^2\int_0^{z_s}\frac{c\,\rmd z}{H(z)}\int \frac{\rmd k_{\perp}}{2\pi}\frac{1}{k_{\perp}^3}P_{\delta}(k_{\perp},z)(1-J_0(k_{\perp}x_{\perp}))\int\frac{\rmd^2\boldsymbol{v}_{\perp}}{2\pi \sigma_v^2}\,{\rm exp}\left(\frac{-v_{\perp}^2}{2\,\sigma_v^2}\right)\,e^{-i\boldsymbol{k}_{\perp}\cdot \boldsymbol{v}_{\perp}\delta t}\\
         &=\frac{8}{c^6}(4\pi G\bar{\rho}_{\rm m})^2\int_0^{z_s}\frac{c\,\rmd z}{H(z)}\int \frac{\rmd k_{\perp}}{2\pi}\frac{1}{k_{\perp}^3}P_{\delta}(k_{\perp},z)(1-J_0(k_{\perp}x_{\perp}))\,e^{-k^2_{\perp}\sigma_v^2\delta t^2/2}.
\end{split}
\end{equation}
The correlation function is exponentially damped for $k_\perp\sigma_v\delta t \gtrsim 1$. However, we can always measure the temporal power spectrum on a shorter timescale to get around with this suppression. 
The temporal power spectrum can be calculated as the Fourier transform of the correlation function 
\begin{equation}
    \begin{split}
        P_t(\omega)&= \int \rmd(\delta t) \, e^{i\,\omega\,\delta t}\,\langle\Delta t(0)\Delta t( \delta t)\rangle \\
        &=\frac{8\sqrt{2\pi}}{c^6}\,(4\pi G\,\bar{\rho}_{\rm m})^2\,\int_0^{z_s}\frac{c\,\rmd z}{H(z)}\int \frac{\rmd k_{\perp}}{2\pi}\frac{1}{k_{\perp}^3}\,P_{\delta}(k_{\perp},z)\,\left(1-J_0(k_{\perp}x_{\perp})\right)\,\frac{1}{k_{\perp}\sigma_v}\,{\rm exp}\left(-\frac{\omega^2}{2\,k^2_{\perp}\,\sigma_v^2}\right).
    \end{split}
    \label{eqn:Ptomega}
\end{equation}
Information about the high-$k$ modes is contained in high-$\omega$ values in the temporal power spectrum. 

Assuming that the errors in timing various bursts behave as white noise with variance $\delta t_m^2$, the uncertainty in estimating the temporal power spectrum is set by $\delta P_t(\omega) =\pi\,\delta t_m^2/(\sqrt{N_m}\omega_0)$ for $\omega < \omega_0$, where $\omega_0=\pi\,N/\delta t$ is the highest frequency determined by the FRB repeating period and $N_m\approx \sigma_v\delta t\, k$ is the number of modes that can be used to constrain the signal. (This expression would be exact if the FRBs from a repeater were periodic.) Since this sensitivity does not depend on $\omega$, we can extract information about the high-$k$ modes from the high frequency part of the temporal power spectrum. By measuring the temporal power spectrum of the arrival time variation, we can expand $J_0$ in Eq.~\ref{eqn:Ptomega} in small $k_\perp x_\perp$, which yields for the leading order the sensitivity to $P_\delta (k)$ per logarithmic $k$ of
\begin{equation}
    \delta P_{\delta}\approx \frac{2\pi\, \delta t_{\rm m}^2/N}{\frac{1}{c^6}(4\pi G\bar{\rho}_{\rm m})^2\,x_0^2\,D_{\rm s}}(\sigma_v \,\delta t\,k)^{1/2}.
\end{equation}
This sensitivity will match the result in Eq.~\ref{eq:PK_t2} when $\sigma_v\,\delta t\, k\sim 1$, but scales differently with $k$. Note that the parameter space of interest satisfies $k_\perp x_\perp \ll 1$ so we can still expand the Bessel function.  This expression only holds for $(\sigma_v \delta t)^{-1} <  k \ll x_0^{-1}$, which is a relatively narrow range for our fiducial $x_0 =20~$AU. (The lower bound of this range is from the finite observation duration, which is not accounted for in our above expressions).  We also note that $x_0/c$ is similar to the tens of hour repetition time for FRBs, which is roughly the maximum wavenumber where the power spectrum can be measured. 
The sensitivity of a $P_t(\omega)$ measurement to the matter power spectrum numerically evaluates to:
\begin{equation}\label{eq:Pk_tempes}
    \frac{\delta P_{\delta}(k)\,k^3} {2\pi^2} \approx 1.1\times 10^4\left(\frac{N}{100}\right)^{-1}  \left(\frac{k}{\rm 100 \,pc^{-1}} \right)^{7/2} \left(\frac{\delta t}{10 \,\rm yr} \right) \left(\frac{\delta t_m}{0.1 \rm ns} \right)^{2}  \left(\frac{D_s}{1 \rm Gpc} \right)^{-1} \left(\frac{x_0}{100\rm AU}\right)^{-2}.
\end{equation}
The sensitivity on the dimensionless matter power spectrum from measuring the temporal power spectrum scales as $k^{7/2}$.


\section{Weak Lensing Effect with Strongly Lensed Fast Radio Bursts}
\label{sec:2_image}

In the previous section, we performed calculations for an observational configuration with two two radio dishes separated on the Solar System scale (the 2-Dish configuration). 
Another configuration that could enable us to achieve the same goal of offset sightlines is the observation of a single repeating FRB source that has multiple images due to strong lensing by astrophysical lenses (such as a galaxy or a galaxy cluster). The rate of such events is expected to be significant with more FRBs detected in the future. The strong lensing probability for FRBs at $z>1$ is $\sim 10^{-4}$ and the apparent repeater fraction of CHIME-discovered FRBs is $2 \times 10^{-2}$ \cite{CHIMEFRB:2023myn}. Assuming these numbers, and if SKA phase 2 achieves $>10^7$ FRBs, as some forecasts suggest \cite{Xiao:2022hkl}, one would expect to detect tens of strongly lensed repeating FRBs per year with the caveat that we need to detect FRBs and FRB repeaters at $z>1$ for a large strong lensing rate and the redshift distribution of repeating FRBs is only weakly constrained (and there is some evidence that the discovered repeaters are from lower redshifts than non-repeating FRBs). Therefore, one would expect to detect tens of strongly lensed repeating sources with years of operation of SKA phase 2. 

As illustrated in the right cartoon of Fig.~\ref{fig:cartoon}, the strongly lensed FRB source will have at least two lensed images, toward which the two light paths probe slightly different regions of the Universe and experience different Shapiro time delays from DM substructures. This situation can be thought of as weak lensing effects superimposed on the strong lensing effects, as DM substructures do not perturb light propagation strong enough to form additional lensed images. The arrival time difference corresponding to the two lensed images can be measured to high precision, just like in the case of the 2-Dish configuration, and one can look for variation in the arrival time differences between the lensed images of the repeated bursts.
Ref.~\cite{Dai2017lensedFRBtiming} explored some astrophysical applications of this effect, but only assumed a conservative FRB timing precision at the millisecond level, which is orders of magnitude worse than the sub-nanosecond timing precision achievable through a time-series correlation analysis.
For the rest of this section, we will discuss the arrival time difference arising from dark matter substructures in strong lensing systems. Then we will discuss possible concerns from decoherence effects on the multiple images.

\subsection{Weak Lensing Effects In Strong Lensing Systems}

Now we study the weak lensing effects from small-scale potential fluctuations on a gravitationally lensed FRB source with two lensed images (2-Image system hereafter). We assume that the radio waves following the different light paths of the two images remain phase coherent and a timing accuracy at the sub-nanosecond level is achievable. The dominant contribution to the arrival time difference between the two images comes from strong lensing, but it is often not possible to model the strong gravitational lens very accurately. Therefore we need to monitor a repeating FRB source to extract useful information about DM substructures. As for the time-varying arrival time difference, the leading-order contribution is essentially what we have calculated in Sec.~\ref{sec:2-Dish}, except that in the 2-Image system there is a much longer transverse ($\sim\,$kpc) baseline between the widely separated light paths corresponding to the two lensed images. The variance of the arrival time difference is simply given by 
\begin{equation}
\begin{split}
     \sigma_t^2   =&\frac{16}{c^6}\,(4\pi G\bar{\rho}_{\rm m})^2\int_0^{z_s}\frac{c\,\rmd z}{H(z)}\int \frac{\rmd k_{\perp}}{2\pi}\frac{1}{k_{\perp}^3}P_{\delta}(k_{\perp},z) \,(1-J_0(k_{\perp} x_{\perp}))\,\left(1-e^{-k^2_{\perp}\sigma_v^2\delta t^2/2} \right)\\
     \approx &\frac{16}{c^6}\,(4\pi G\bar{\rho}_{\rm m})^2\int_0^{z_s}\frac{c\,\rmd z}{H(z)}\int \frac{\rmd k_{\perp}}{2\pi}\frac{1}{k_{\perp}^3}P_{\delta}(k_{\perp},z) \,\left(1-e^{-k^2_{\perp}\sigma_v^2\delta t^2/2} \right)\\
      \approx&\frac{16}{c^6}\,(4\pi G\bar{\rho}_{\rm m})^2\int_0^{z_s}\frac{c\,\rmd z}{H(z)}\int \frac{\rmd k_{\perp}}{2\pi}\frac{1}{k_{\perp}^3}P_{\delta}(k_{\perp},z)\left( \frac{k^2_{\perp}\sigma_v^2\delta t^2}{2}-\frac{k^4_{\perp}\sigma_v^4\delta t^4}{8}+ ... \right),
\end{split}
\end{equation}
where the second line approximates $1-J_0(k_\perp x_\perp)\approx 1$, as for widely separated sightlines $k_\perp x_\perp \gg 1$. Therefore the variance of the arrival time difference has no dependence on the image separation. The reason behind this is that the DM substructures intervening the two light paths will be uncorrelated. The temporal variation of the arrival time difference simply comes from the motion of DM substructures near the light paths. It is worth noting that the linear term on $\delta t^2$ will be more sensitive to DM structures at larger scales while the quadratic term has the same behavior as the 2-Dish system with a $\sim \sigma_v\delta t$ (1000 AU for the fiducial value) baseline. To see this, we apply the same calculation for the second derivative in the 2-Image scenario, and take the long baseline limit
\begin{equation}
    \sigma_{t^{\prime\prime}}^2
      =\frac{24}{c^6}(4\pi G\bar{\rho}_{\rm m})^2\sigma_v^4\int_0^{z_s}\frac{c\,\rmd z}{H(z)}\int \frac{\rmd k_{\perp}}{2\pi}\,k_{\perp}\,P_{\delta}(k_{\perp},z).
\end{equation}
This leads to a sensitivity on the matter power spectrum
\begin{equation}\label{eq:PK_2i}
    \delta P_\delta(k)\approx \frac{2\pi\, \delta t_{\rm m}^2/N}{\frac{24}{c^6}(4\pi G\bar{\rho}_{\rm m})^2\,D_{\rm s}\,\sigma_v^4 \,\delta t^4\,{k^2}}.
\end{equation}
The above expression can be compared to the sensitivity coming from the first time derivative of the arrival time variation for the 2-Dish system, as discussed in Eq.~\ref{eq:Pk_t1}.
The sensitivity of the 2-Image configuration is equivalent to a 2-Dish configuration with a baseline of $2000 \,{\rm AU}\,(\delta t/10\,{\rm yr})\,(\sigma_v/10^3\,{\rm km}{\rm s}^{-1})$. Note that the second time derivative term in the 2-Image scenario has the same $k$ scaling as the first derivative term in the 2-Dish system. Likewise, when we expand the 2-Image signal to the third time derivative (cubic term), it has the same $k$ scaling as the second time derivative term discussed in Eq.~\ref{eq:PK_t2}. The difference in the $k$ scaling, for $k\,x_0 < 1$, results from the difference that the two slightlines to the two receiver dishes probe similar potential fluctuations while in the strong lensing scenario potential fluctuations near the two widely-separated light paths are uncorrelated.
Lower time derivatives of the arrival time difference are more contaminated by long wavelength modes in the $\Lambda$CDM scenario. This limitation can be circumvented by measuring higher-order time derivatives of the arrival time difference.

It might seem like the strong lensing scenario is superior in sensitivity and there is no need for VLBI in space. 
However, as we will discuss later in this section, in the strong lensing scenario the signal is likely to be more prone to the decoherence effect or scattering effect, depending on the properties of the FRB emission sources. These effects may substantially degrade the sensitivity in the strong lensing scenario.

There is also an effect in the arrival time at second order in DM potential fluctuations. 
Due to DM potential fluctuations everywhere along the path, the otherwise straight light paths from the source to the strong lensing plane and from that plane to the observer are not exactly extremized light paths and hence are perturbed.
In Appendix.~\ref{app:sec_order}, we estimate that such correction to the arrival time can provide a signal comparable to a 5 AU baseline, which is subdominant.

\subsection{Decoherence and Scattering Effects Between Sightlines}

Realizing the method we propose requires that the FRB radio signals that travel along different sightlines have phase coherent voltage time series to allow cross-correlation. 
We now examine the plausibility of this assumption for both the 2-Dish configuration and the 2-Image scenario. The most outstanding issue of phase decoherence likely comes from the interstellar medium (ISM) of the Milky Way.
ISM can cause pulse delay and broadening along sightlines to individual radio receivers in space. This is known as the scattering effect, which is induced by density inhomogeneities in the interstellar plasma. Observations of pulsars and FRBs away from the Galactic disk suggest a scattering time $\delta t_{\rm sc} \simeq 30\, {\rm ns} \,(\nu/\rm GHz)^{-4}\,(\sin b)^{-2.5}$, where $b$ is the Galactic latitude of the sightline \citep[e.g.][]{Boone:2022pdz}. Pulse delay and broadening are typically on this timescale. If the scattering-induced time delay is significant ($\delta t_{\rm sc}>1/\nu$), radio waves travel along multiple geometric paths with significant phase differences, resulting in decoherence in the voltage time series. To have $\delta t_{\rm sc} < \nu^{-1}$, observations need to be carried out at $\nu \gtrsim3$ GHz to suppress ISM decoherence. For our fiducial $0.1\,$ns timing accuracy, we would require $\delta t_{\rm sc} < 0.1\,{\rm ns}$; otherwise the timing error worsens by scattering. See Ref.~\cite{Boone:2022pdz} for more extensive discussions on the scattering effects on FRB observations in the context of space interferometry on the Solar System scale.  Finally, scattering in the lens itself could be important since the light paths may pass close to the denser lens center and its host galaxy's ISM.


Another requirement for phase coherence is that the FRB source must remain unresolved to the baseline spanned by the two sightlines. This requires the source size to satisfy $l_{\rm FRB}< \lambda/\theta_E$, where $\theta_E$ is the angular separation of two sightlines with respect to the source and $\lambda$ is the wavelength of the radio wave. In the case of the 2-Dish configuration, we expect an angular separation as small as $\theta_E = x_0/D_s \sim 10^{-12}$, and so  $l_{\rm FRB} > 10^{13}\rm cm$ which should be easily satisfied \citep{Boone:2022pdz}. The scattering effect in the host galaxy may make the effective size of the emission region larger but again this is a sufficiently small effect for the 2-Dish system for even the most host-galaxy-scattered FRBs that have been discovered \cite{Boone:2022pdz}. 

The requirements are more strigent for the case of strongly lensed FRBs for FRB timing:  The angular separation of the two images must be $\theta_E\sim 10^{-6}\,$rad to not resolve the source, requiring the FRB source size to be smaller than $\sim 10^7$ cm if observed in the gigahertz range.  The size requirement can be satisfied if each FRB event has a source comparable to or more compact than the size of a neutron star ($\sim$10 km). Since the size of the FRB emission region is not fully understood, this condition may be violated in other models and we would not have two strongly lensed images that remain coherent with each other. Therefore the determination of the FRB emission mechanism will confirm if lensed FRBs can remain coherent over two paths. See Ref.~\cite{Xiao:2022hkl} for a similar discussion on the decoherence of strongly lensed FRB images from cosmic strings. Even if the emission spot of each FRB burst is sufficiently small, there may be spatial variation in the exact location of the emission spot relative to the hosting compact object (e.g. if FRB bursts are sourced by a neutron star) from burst to burst. See Ref.~\cite{Dai2017lensedFRBtiming} for a discussion on random variations in the arrival time difference due to such spatial variation in the emission spot location of individual bursts in the strong lensing situation. For typical strong lensing parameters, random spatial variation of the emission spot over distances $\gtrsim 100\,$km can produce $\gtrsim\,$ns random noises in the arrival time difference, which may be relevant for FRBs from a neutron star. 
Even if the emission size of FRB source is small, the effective image size will be larger with multi-path diffractive propagation from electron inhomogeneities in the host galaxy, which imposes an additional requirement on how unscattered the FRB source needs to be. See Ref.~\cite{Leung:2022vcx} for discussions on the decoherence effect in the host galaxy with a scattering screen model. The scattering effect in the host galaxy needs to be suppressed to avoid a large effective emission region by either observing at high frequencies ($\gtrsim 5$ GHz) or looking at unscattered FRBs.  Some FRBs may reside in galaxies that are evacuated of their ISMs or be far enough in the galactic outskirts to be essentially unscattered by their host galaxy. 
There are some FRBs where the limits on scattering are smaller than a microsecond at 1 GHz \cite{Nimmo:2020sva,Cho:2020gtg}. 
When the delay caused by the scattering screen in the host galaxy is less than $1/\nu$, it will behave refractively and will not substantially enlarge the image.  If the scattering delay is sufficiently small, the lens will not resolve the source, and the FRB can be treated as the same light that travels through the host galaxy with multiple paths and hence would remain coherent through the host galaxy over the lensed pathways.

Another potentially confounding effect is the microlensing effects from stars intervening the sightlines, which however should be negligible as discussed in Ref.~\cite{Xiao:2022hkl}. For solar mass objects like stars, their impact parameters to the sightlines are much greater than their size. Therefore the main effect comes from their mass but not their compactness. On average, the fractional cosmic matter density in stars is small compared to that in DM. 


\section{Physical Applications to Enhanced Dark Matter Substructures}
\label{sec:application}

In the previous Sections, we have shown that precision timing of repeating FRBs can be used to probe the matter power spectrum on very small scales, realized either with a very long interferometry baseline of the Solar System size or with a strongly lensed FRB source with multiple lensed images. The matter power spectrum on scales $k> 10\,{\rm Mpc}^{-1}$ is only weakly constrained by existing observations.  
Matter clustering in the standard $\Lambda$CDM cosmology is usually below the detection threshold at high $k$ due to the lack of dense structures on small scales. On the other hand, there are well-motivated models of nonstandard early Universe dynamics in which the formation of DM substructures on minuscule scales is enhanced. Measuring the matter power spectrum on such scales will be a strong test of the $\Lambda$CDM cosmology, and provides new discovery opportunities for new physics about the nature of the DM and the thermal history of the very early Universe. Models featuring enhanced matter power spectrum on minuscule scales usually involve interesting cosmological dynamics in the early Universe because the size of the self-gravitating structures formed is related to the horizon length at early times when nontrivial dynamics occurred. Axion DM from a post-inflationary Peccei-Quinn phase transition, early matter domination with a low reheating temperature, and vector DM produced during inflation are the most important model examples that give rise to the formation of DM substructures on small scales. In the following, we apply the calculations we have performed in previous sections to these models to depict the future discovery opportunity with precision FRB timing.

\subsection{Axion Miniclusters}

The axion particle is a solution to the strong CP problem and may account for the DM~\cite{Peccei:1977hh,Weinberg:1977ma,Wilczek:1977pj,Kim:1979if,Abbott:1982af,Dine:1982ah,Preskill:1982cy, Peccei:2006as}. Currently the allowed parameter space of axion is broad, and axions not intended to solve the strong CP problem are referred to as axion-like particles, which have broader parameter space. If the Peccei-Quinn symmetry breaking is after inflation, the axion field will take different values in different horizon patches before the critical time when the axion acquires its mass. The inhomogeneity in the axion field value will convert to matter density fluctuations since there is an energy transfer from the QCD vacuum to axion matter density, leading to the formation of axion miniclusters when axion dark matter starts to dominate the energy density budget of the Universe \cite{Kolb:1993zz,Hogan:1988mp,Kolb:1993hw}. This production mechanism of axion relic density is known as the vacuum misalignment mechanism. During the evolution of axion fields after the symmetry breaking, topological defects such as axion strings will also form in the post-inflationary scenario, which will eventually dissipate and contribute to the axion relic density \cite{Vaquero:2018tib,Buschmann:2019icd,Buschmann:2021sdq,Gorghetto:2020qws}. This will also extend the matter power spectrum expected from the vacuum misalignment mechanism to subhorizon scales due to the structure of strings, leading to a smaller characteristic minicluster mass \cite{Vaquero:2018tib,Pierobon:2023ozb}.
There are other scenarios of axion physics that can also lead to enhanced substructures on small scales, such as the kinetic misalignment mechanism \cite{Barman:2021rdr,Eroncel:2022efc,Chatrchyan:2023cmz}, the large misalignment mechanism \cite{Arvanitaki:2019rax}, a nonstandard thermal history \cite{Nelson:2018via,Visinelli:2018wza}, and inflated domain walls that re-enter the horizon \cite{Harigaya:2022pjd,Redi:2022llj,Gorghetto:2023vqu}. 

The cosmological evolution of axion miniclusters in the post-inflationary scenario has been studied using semi-analytic techniques \cite{Enander:2017ogx,Blinov:2019jqc,Fairbairn:2017sil,Ellis:2020gtq,Ellis:2022grh} and N-body simulations \cite{Zurek:2006sy,Eggemeier:2019khm,Xiao:2021nkb,Eggemeier:2022hqa,Shen:2022ltx}, which suggest that an order unity fraction of the axion DM will be locked in axion miniclusters, motivating astrophysical searches for those DM substructures.
In the following, we will focus on the vacuum misalignment mechanism that results in a white-noise matter power spectrum at the scale of horizon size when axions obtain mass from their potential. The linear matter overdensity has a characteristic variance on a scale corresponding to a comoving wavenumber $k$:
\begin{equation}
    \Delta^2(k,z) = D^2_1(z)\,A_{\rm osc} \left(\frac{k}{k_{\rm osc}}\right)^3, \; k<k_{\rm osc}
\end{equation}
where $D_1(z)$ is the growth function normalized at matter-radiation equality and $A_{\rm osc} \sim \mathcal{O}(1)$ is the amplitude of the white-noise power spectrum, which is taken to be $A_{\rm osc}=1$ in this work. The exact value of $A_{\rm osc}$ is degenerate with $k_{\rm osc}$ since the initial minicluster mass is the only relevant parameter. The size of the density fluctuation is expected to fall off fast at $k>k_{\rm osc}$. Simulations of axion string dynamics show that strings will contribute significantly to the axion density fluctuations, leading to a smaller $A_{\rm osc}$ and a cutoff at high $k$, which effectively lowers the mass of initial axion miniclusters. For the QCD axion, $k_{\rm osc}$ can be related to the axion mass $m_a$ \cite{Sikivie:2006ni,Vaquero:2018tib}
\begin{equation}
    \frac{1}{k_{\rm osc}} = 0.036\,{\rm pc}\,\left(\frac{50 \,\rm \mu eV}{m_a}\right)^{0.17},
\end{equation}
Where the numerical power-law index accounts for the temperature dependence of the axion mass during the establishment of the axion relic abundance around the time of QCD phase transition. Therefore, the comoving scale corresponding to the QCD axion is extremely small, but the growth of density fluctuations on these scales and the later formation of bound structures do result in potential fluctuations even on the AU scales that our proposed method can probe. 

We estimate the nonlinear matter power spectrum predicted in the axion DM model and compare it to the reach of our proposed method as discussed in previous Sections. Since the dimensionless matter power spectrum almost starts with nonlinear values at matter-radiation equality, we adopt the halo model to calculate the nonlinear matter power spectrum. As discussed in Appendix.~\ref{app:halo_model}, the halo model extends the nonlinear matter power spectrum further to even smaller scales by accounting for the density profiles of the axion miniclusters. The axion miniclusters are expected to be incorporated into the larger DM structures, so we integrate over the $\Lambda$CDM halo mass function and populate the large halos with axion miniclusters as subhalos. According to the halo model, the 1-halo term from the axion miniclusters inside large halos is the dominant contribution to the nonlinear power spectrum on small scales. 
In the halo model, the density field is expressed in terms of halos as building blocks. The Fourier transform of the halo density profile weighted by the halo mass function gives the correlation function of the density field and thus the nonlinear matter power spectrum.

\begin{figure}[h!]
    \centering
\includegraphics[width=12cm]{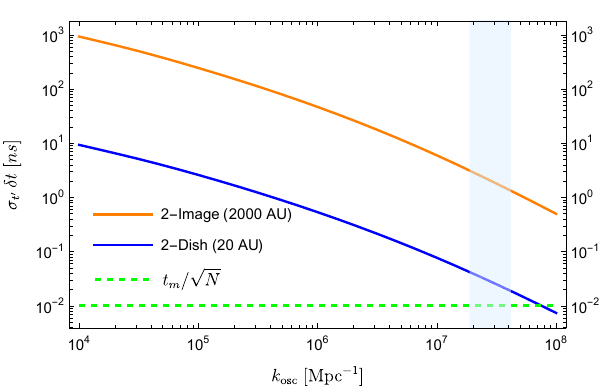}
\caption{Variance of the first time derivative of the arrival time difference for the 2-Dish configuration and 2-Image system (which is equivalent to a 2000 AU baseline) as calculated in Eq.~\ref{eq:sigmat'} as a function of $k_{\rm osc}$, which is the comoving Hubble scale at the time when the axion acquires its mass. The blue band corresponds to the axion mass range from 5 $\mu$eV to 500 $\mu$eV, which is the estimated mass range that leads to the correct DM relic abundance in the scenario of the QCD axion with a Peccei-Quinn phase transition after inflation. The green dashed curve is $\delta t_m/\sqrt{N}$, which is noise floor for a timing precision of $t_m=0.1$\,ns per measurement and $N=100$ measurements. The fiducial value we take for the dish separation is $x_0=20\,$AU. }
    \label{fig:FF}
\end{figure}

We now discuss the scales that contribute most to the detectable signal. In Eq.~\ref{eq:sigmat'}, we have shown that $\sigma_{t^{\prime}} \propto \int \rmd k \,k\, P_\delta(k) $. Therefore, the integral is sensitive to a wide range of scales at $k>k_{\rm osc}$ for which our modeling finds that $P_\delta(k)\propto k^{-2}$, as shown in Fig.~\ref{fig:power}. This $k^{-2}$ scaling owns to the density profile of halos and their mass function.
Therefore we expect the 2-Dish system or the 2-Image system will mostly probe the high $k$ tail ($k\gtrsim k_{\rm osc}$) of the matter power spectrum enhanced by axion miniclusters. 
\begin{figure}[h!]
    \centering
\includegraphics[width=15cm]{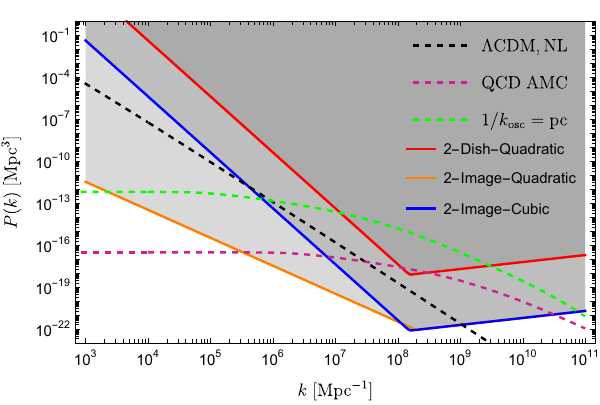}
\caption{The sensitivity of measuring the time variation of the arrival time difference of repeating FRBs with the 2-Dish configuration or in the 2-Image scenario. In the 2-Image scenario, the repeating FRB source is strongly lensed with multiple images. The nonlinear matter power spectrum in the standard $\Lambda$CDM cosmology is plotted as the black dashed curve. The 2-Dish sensitivity is calculated assuming a dish separation of $20\,$AU. Matter power spectrum above the orange curve can be probed with the 2-Image scenario, while the matter power spectrum above the red curve can be probed with the 2-Dish configuration using the second-order time derivative (2-Dish-Quadratic) of the arrival time difference. Note that the cubic term in the 2-Image system is equivalent to the quadratic term in the 2-Dish case.
The magenta and green dashed curves are the matter power spectrum arising from axion miniclusters (AMC) and the magenta dashed curve is that expected from the QCD axion. The green dashed curve has a larger axion minicluster mass than the QCD axion minicluster, which is more detectable.
We found that the QCD axion miniclusters will be detectable by the 2-Dish configuration with a $20\,$AU baseline.}
\label{fig:Pk_amc}
\end{figure}
In Fig.~\ref{fig:FF}, the variance of the first time derivative of the arrival time difference is plotted as a function of $k_{\rm osc}$. 
Results for both the 2-Dish configuration and the 2-Image scenario are plotted. For a smaller $k_{\rm osc}$, the axion miniclusters have larger masses and are more detectable. As can be seen in Fig.~\ref{fig:FF}, the 2-Dish configuration with a $20\,$AU baseline is sufficiently sensitive for QCD axion miniclusters of a wide range of masses. In other words, the nonlinear power spectrum calculated from axion miniclusters is within the reach of FRB timing for the whole range of QCD axion parameters. 

In Fig.~\ref{fig:Pk_amc}, we plot the sensitivity to the nonlinear matter power spectrum and compare it to the expection from axion miniclusters. As shown in Fig.~\ref{fig:Pk_amc}, the high-$k$ portion of the nonlinear matter power spectrum from axion miniclusters is above the detection limit. This is in agreement with our previous discussion that the $k\gtrsim k_{\rm osc}$ portion of the matter power spectrum has the dominant contribution to the signal. It is worth noting that axion miniclusters may have totally different masses if our consideration is not restricted to the QCD axion. An axion-like particle that starts to oscillate at later times will correspond to a smaller $k_{\rm osc}$ and thus more massive miniclusters. Those models will be more detectable in general. As an example, in Fig.~\ref{fig:Pk_amc} we show the result for a slightly larger minicluster mass, $M_0=(4/3)\pi\bar{\rho}_m(1/k_{\rm osc}^3)=1.67\times 10^{-7}M_{\odot}$, or $k_{\rm osc}=1\,\rm pc^{-1}$, with the green dashed curve. This scenario is significantly above the detection threshold of the 2-Dish configuration. 

The sensitivity presented here can be alternatively applied to the scenario of the DM comprised of asteroid to planet mass primordial black holes (PBHs), which source a white-noise initial matter power spectrum on length scales larger than the inter-PBH separation, the same as in the case of axion DM with a Peccei-Quinn phase transition after inflation. The PBH mass plays the role of the axion minihalo mass and PBHs become gravitationally bound in PBH clusters. However, PBH clusters could also have contributions at very high $k$ values from lensing of individual PBHs in the cluster, which can potentially enhance observable effects. For example, PBHs with $M\sim 10^{-13}M_{\odot}$ are currently not constrained by microlensing and might be detectable with the 2-Dish configuration. 
The 2-Image system, if the two images remain coherent, is equivalent to a 2-Dish system with a $2000\,$AU baseline and can be even more sensitive to PBH clusters. The 2-Image scenario may be so sensitive that the $\Lambda$CDM halos will have a significant contribution to observables of low-order time derivatives. Similar to the 2-Dish configuration, we can use higher-order time derivatives to filter out contributions from low $k$ modes. In any case, the non-standard minuscule DM substructures are still detectable as long as they dominate the signal. The confounding signal from the standard $\Lambda$CDM halos is only a concern when we want to probe the matter power spectrum way below the predicted signal strengths from QCD axion miniclusters.

Currently the most stringent constraint on the isocurvature initial matter power spectrum from axion miniclusters or PBH clusters comes from ly$\alpha$ forest, which requires $k_{\rm osc}>1600\,{\rm Mpc}^{-1}$ \cite{Irsic:2019iff}. 
FRB interferometry will be orders of magnitude more sensitive to minuscule DM substructures. Another advantage of our probe is that DM substructures are more significantly disrupted inside galaxies due to frequent gravitational encounters with passing stars \cite{Shen:2022ltx} while the intergalactic substructures we study here can evade dynamic destruction by stars in galaxies.


\subsection{Early Matter Domination}

The standard $\Lambda$CDM model of cosmology has been a great success in explaining current observations, ranging from Big Bang Nucleosynthesis (BBN) at early times to large scale structure formation at late times. Inflation solves the horizon and flatness problems of the hot Big Bang cosmology and generates curvature perturbations as the seeds for large-scale structure formation. However, a remaining obscure component of the physical cosmology model is the reheating process with the thermal history before BBN, which evades all current observations. The most stringent bound on the maximal temperature of radiation domination comes from the thermal production of neutrinos. If the Universe is radiation-dominated at a temperature of $\sim 3\,$MeV or higher, thermal neutrino production is sufficient to produce the correct abundances of light elements \cite{Kawasaki:1999na,Kawasaki:2000en,Hannestad:2004px,Ichikawa:2005vw}. There are no observational constraints on the thermal history at temperatures above $3\,$MeV.

On the other hand, nonstandard thermal history such as an early epoch of matter domination or a first-order phase transition is required to explain the asymmetry between baryon and anti-baryon, which is an outstanding puzzle in cosmology. It is also possible to generate the baryon asymmetry at a temperature of MeV scale with meson decays in an epoch of early matter domination \cite{Elor:2018twp,Nelson:2019fln,Alonso-Alvarez:2019fym,Elor:2020tkc,Elahi:2021jia}.
\begin{figure}[h!]
    \centering
\includegraphics[width=15cm]{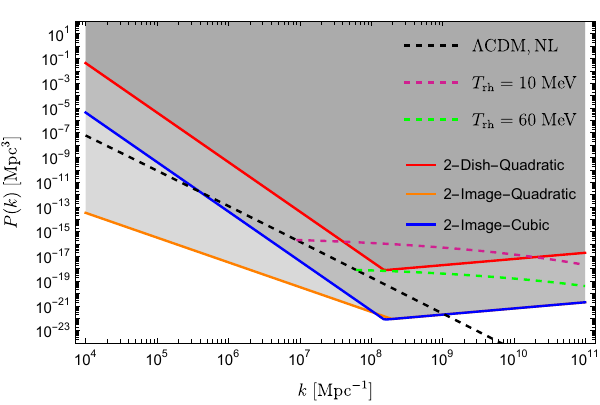}
\caption{ The sensitivity of various probes on the matter power spectrum from early matter domination. All the sensitivity curves are the same with Fig.~\ref{fig:Pk_amc}. Here we presented two more curves that correspond to the nonlinear power spectrum of dark matter minihalos from early matter domination. The reheating temperatures we considered are $10\,$MeV and $60\,$MeV. Higher reheating temperatures will produce lighter dark matter minihalos, which are more challenging to detect. The green dashed curve, corresponding to a reheating temperature of $60\,$MeV, is within the reach of the 2-Dish system with a 20 AU baseline. }
    \label{fig:Pk_constraint_EMD}
\end{figure}
Since the energy density of matter scales as $a^{-3}$ while that of radiation scales as $a^{-4}$, matter will always dominate the energy density budget as long as it is stable. In the early Universe, if there are long-lived scalar fields such as stabilized moduli in string theories \cite{Mollerach:1989hu,Lyth:2001nq,Moroi:2001ct}, those scalar fields can dominate the energy density before they decay. It can be perfectly consistent with current constraints from the light elements abundances produced during BBN as long as those scalar fields reheat the Universe to a temperature above $3\,$MeV. 

Such nonstandard thermal history will have unique effects on structure growth. During matter domination, subhorizon modes of matter density fluctuations grow linearly with the scalar factor. For modes that enter the horizon during the early matter domination ($k \gtrsim a_{\rm rh} H_{\rm rh}$), the growth of matter perturbation after early matter domination can be written as \cite{Erickcek:2011us} 
\begin{equation}\label{eq:D_emd}
    \delta_{\rm dm}\approx \frac{2}{3}\frac{k^2}{a_{\rm rh}^2H_{\rm rh}^2}\Phi_0,
\end{equation}
Where $a_{\rm rh}$ and $H_{\rm rh}$ are the scale factor and Hubble at reheating respectively, which is the end of early matter domination.
$\Phi_0$ is the initial amplitude of potential perturbation before the onset of early matter domination. Since we discuss adiabatic fluctuations that are amplified during early matter domination, the matter density perturbations will follow the potential perturbation.
A larger value of $k$ corresponds to earlier horizon entry and thus more linear growth. If $k\lesssim a_{\rm rh} H_{\rm rh} $, those modes remain superhorizon and will not grow during early matter domination.
The scalar factor at horizon entry scales as $a_H\propto k^{-2}$ because the Hubble scales as $H\propto a^{-3/2}$ during matter domination and therefore the ratio of momentum to horizon scale $k/(aH)$ will behave as $k\, a^{1/2}$. This justifies the $k^2$ dependence on the growth function in Eq.~\ref{eq:D_emd}. Note that we still need the transfer function in the late Universe to extrapolate the result from the early Universe to the current Universe.

In Appendix.~\ref{app:halo_model}, we discuss the mass function of DM minihalos produced from an early epoch of matter domination and the corresponding nonlinear matter power spectrum. In Fig.~\ref{fig:Pk_constraint_EMD}, we present the sensitivity of our proposed method to the matter power spectrum as calculated in previous Sections as well as the nonlinear matter power spectrum due to DM minihalos from the early matter domination. As shown in Fig.~\ref{fig:Pk_constraint_EMD}, the 2-Dish system with a separation of $20\,$AU will be able to probe early matter domination with a reheating temperature of 60 MeV, which is above the lowest allowed value of reheating temperature, $\sim 3\,$MeV. The 2-Image system, which only requires the observation of one strongly lensed FRB repeater, can potentially probe early matter domination up to a reheating temperature of $600\,$MeV. For such high reheating temperatures, contributions from $\Lambda$CDM halos might dominate the signal. To improve, we can study higher-order time derivatives or the temporal power spectrum of the arrival time difference, which is sensitive to higher-$k$ modes.
Therefore, we may learn more about the thermal history of the early Universe before BBN by measuring the gravitational effects of minuscule DM substructures.

\section{Conclusion}
\label{sec:conclusion}

In this work, we have proposed a new probe of DM substructures using a precision comparison of the arrival times of repeated FRB radio signals along two offset sightlines.  Owing to gravitational time delays and the cosmological motion of minihalos -- moving a hundred AU per year in projection --, as long as there are structures on such scales there should be slight arrival time variations that we showed may be detectable with coherent timing of the FRB signals.  We have shown that the new probe is sensitive to a population of collapsed minuscule DM structures from subgalactic masses all the way down to $10^{-13}\,M_{\odot}$. The multiple sightlines toward a single FRB source may be realized with a constellation of radio receivers in space separated by $0.1-100\,$AU or in a strong lensing situation with a single receiver recording the FRB signals from multiple lensed images. We have calculated the arrival time difference and have quantified the temporal variation that is gravitationally induced by intervening DM substructures.   

Our first proposal involves performing radio interferometry with baselines on the solar system scale, which requires sending radio dishes to deep space with separations of at least tens of AU as required to start constraining the most motivated QCD axion and EMD models. The position of the dishes would have to be accurately calibrated using three other receivers using trilaterations -- this ambitious mission would also measure the parallax distance to FRBs and hence the cosmic expansion history, with tens of AU baselines potentially providing sub-percent distance constraints \cite{Boone:2022pdz}. We have shown that there will be a simultaneous interesting science case for such a mission.  Measuring the linear variation of arrival time differences between two dishes is sensitive to the larger halos present in the vanilla $\Lambda$CDM model, but would also rule out larger parsec to AU-scale enhancements in the matter power spectrum.  Measuring the quadratic change of these time differences (or even the power spectrum of these differences) isolates the parsec-scale and smaller enhancements, even reaching the sensitivity required for motivated QCD axion and EMD models.  A related technique, which is somewhat less sensitive but that does not require repetitions, is to use the geometry of arrival time differences from a constellation of $\geq 4$ receivers separated by tens of AU.  We further showed that the magnifications of type 1a supernova is sensitive to the same signal as our linear signal (and with the sensitivity of a $0.14\,$AU baseline)  -- using existing limits on the magnification of these sources to place new constraints on the matter power spectrum.

Our second approach uses strongly lensed repeating FRB sources, which connect to the observer via multiple separated light paths. Finding strongly lensed repeating FRBs is hopeful with the increased sensitivity of the upcoming radio surveys and a rapidly growing catalog of detected FRB sources. Future surveys like SKA may detect tens of strongly lensed FRBs each year and some of these may eventually be found to repeat. We have shown that observing two lensed images of a single source enables a sensitivity to minuscule DM structures similar to that of radio dishes with a baseline of 2000 AU. While not requiring solar system-scale baselines, the challenges to this method are that a lensed repeating FRB source would need to be discovered in the first place and precision subtraction of atmospheric delays (or going to space) would be required. For both approaches, radio wave coherence along the two disparate paths through the Milky Way's ISM requires observing at $\nu\gtrsim 3$ GHz, but the strong lensing approach would be further challenged by ISM scattering decoherence within the lens. The lensing scenario also requires the location of the FRB emission site to not vary relative to the neutron star's position to $\lesssim 100$ km.

Both approaches would be sensitive to DM substructures even at scales $k\sim 10^{8}\,\rm Mpc^{-1}$, and the 2-Dish system with a separation of $20\,$AU can eventually probe miniclusters in the QCD axion DM model (with initial mass $\sim 10^{-13}M_{\odot}$). DM minihalos produced from an epoch of early matter domination with a reheating temperature $<$60 MeV can also potentially be probed with this 2-Dish configuration, providing a unique gravitational probe to the cosmic thermal history before BBN. More aggressive proposals such as a $1000\,$AU baseline or the strongly lensed FRB repeaters can even probe a reheating temperature of $500\,$MeV.
Primordial black hole (PBH) DM with masses of $\sim 10^{-13}M_{\odot}$ would also be probed by our method with a 20 AU baseline in the 2-Dish system since they produce a very similar white-noise matter power spectrum as axion miniclusters and in this mass range PBH DM is currently not ruled out by microlensing \cite{Green:2020jor}. Longer baselines or strongly lensed FRBs can probe even lighter PBH dark matter, leading to better sensitivities to the most challenging mass range of PBH dark matter.


Compared to two other proposals that probe the small-scale matter power spectrum, pulsar timing arrays \cite{Lee:2020wfn} and astrometry of Galactic stars \cite{Dror:2019twh, VanTilburg:2018ykj}, our method has the potential to provide the most stringent constraints on extremely small scales ($k\gtrsim 10^{8}\,\rm Mpc^{-1}$), which owes to the advantage of observing FRB sources at cosmological distances and the superior timing precision possible with radio interferometry. In future works, we plan to further compare with these proposals and with the proposal of using extreme magnification of extragalactic lensed stars in Ref.~\cite{Dai:2019lud}. 


\section*{Acknowledgement}

The authors thank Tanner Trickle, Wei Xue, Masha Baryakhtar, Junwu Huang, Ken Van Tilburg, and Neal Weiner for useful discussions. HX is supported by Fermi Research Alliance, LLC under Contract DE-AC02-07CH11359 with the U.S. Department of Energy. This work was performed in part at the Aspen Center for Physics, which is supported by National Science Foundation grant PHY-2210452. LD acknowledges research grant support from the Alfred P. Sloan Foundation (Award Number FG-2021-16495), and support of Frank and Karen Dabby STEM Fund in the Society of Hellman Fellows. MM acknowledges support from NSF award AST-2007012 and NASA  NIAC Phase I award 23-NIAC24-B-0029.

\appendix

\section{Path Deflection on 2-Image Timing}
\label{app:sec_order}

Aside from the major effect we discussed in the main text about arrival time difference in the 2-Image system, there will be secondary effects because of the path deflection induced by spatial variance of weak lensing effects. In this section, we will quantitatively calculate the path deflection caused by dark matter substructures and show this effect is subdominant.
According to Fermat's principle, weak lensing effects from gravitational potential fluctuations caused by DM substructures, on top of strong lensing, perturb the light path in such a way the light travel time is extremized.
We first revisit the standard formalism of strong lensing, before we treat the additional weak lensing effect. The time delay is the sum of geometric and Shapiro terms:
\begin{equation}
    t(\boldsymbol{\theta}) = \frac{1+z_d}{c}\frac{d_d\,d_s}{d_{ds}}\left(\frac{1}{2}(\boldsymbol{\theta}-\boldsymbol{\beta})^2-\psi(\boldsymbol{\theta})\right),
\end{equation}
where $\psi$ is the lensing potential and $z_d$ is the redshift of the strong lensing plane. The image and source angular positions are denoted by $\boldsymbol{\theta}$ and $\boldsymbol{\beta}$, respectively. We define $d_d$, $d_s$ and $d_{ds}$ to be angular diameter distances from the Earth to the lens, from the Earth to the source and from the lens to the source, respectively. These should not be confused with the comoving distances involved in discussions in previous Sections.
The ray equation is obtained by finding paths of extremized travel time following Fermat's principle:
\begin{equation}\label{eq:lens_o}
   (\boldsymbol{\theta}-\boldsymbol{\beta})-\boldsymbol{\nabla}_{\theta}\psi=0.
\end{equation}
We treat the weak lensing effect from DM substructures as a perturbation to strong lensing. It is straightforward to include the extra Shapiro time delays, $t_w(\boldsymbol{\theta})$, due to DM substructures and study the resultant change of the light path. The new time delay is thus written as
\begin{equation}
    t(\boldsymbol{\theta}) = \frac{1+z_d}{c}\frac{d_d d_s}{d_{ds}}\left(\frac{1}{2}(\boldsymbol{\theta}-\boldsymbol{\beta})^2-\psi(\boldsymbol{\theta})\right)+t_w(\boldsymbol{\theta}),
\end{equation}
We approximate that the small potential fluctuations induced by the DM substructures do not lead to substantial curvature of the light path away from the strong lensing plane. The gradient of $t_w(\boldsymbol{\theta})$ can be calculated in the same way as we studied for the 2-Dish configuration since it is directly related to the gradient field of the gravitational potential, except that the straight light path intersects the strong lensing plane at a revised location. The new ray equation is obtained by including the spatial derivative of $t_w(\boldsymbol{\theta})$
\begin{equation}\label{eq:lens_new}
     \frac{1+z_d}{c}\frac{d_d\,d_s}{d_{ds}}\left((\boldsymbol{\theta^{'}} -\boldsymbol{\beta})-\boldsymbol{\nabla}_{\theta}\psi\right)+\boldsymbol{\nabla}_{\theta}t_w=0,
\end{equation}
where $\boldsymbol{\theta}^{'}$ represents the perturbed light path intersecting the strong lensing plane after the effect of DM substructures is included. The path change, $\boldsymbol{\theta}^{'} - \boldsymbol{\theta}$, satisfies the equation
\begin{equation}
\label{eq:dtheta}
    \left((\boldsymbol{\theta^{'}}-\boldsymbol{\theta} )-\underbrace{(\boldsymbol{\theta^{'}}-\boldsymbol{\theta} )\cdot \boldsymbol{\nabla}_{\theta}\boldsymbol{\nabla}_{\theta}\psi}_{\rm profile\, dependent}\right) \frac{1+z_d}{c}\frac{d_d\,d_s}{d_{ds}}=-\boldsymbol{\nabla}_{\theta}t_w.
\end{equation}
The exact change depends on the profile of the strong lens, but we will later express the major conclusion in a lens-independent way.
The above equation can be recast into the form
\begin{equation}
    \boldsymbol{\theta^{'}}-\boldsymbol{\theta} = - \left( \frac{1+z_d}{c}\,\frac{d_d\,d_s}{d_{ds}}\right)^{-1}\, \boldsymbol{A}^{-1}(\boldsymbol{\theta})\cdot \boldsymbol{\nabla}_\theta t_w,
\end{equation}
where $\boldsymbol{A}(\boldsymbol{\theta})$ is the $2\times 2$ strong lensing ray deformation matrix and has components $A_{ij}(\boldsymbol{\theta}) = \delta_ij - \nabla_i \nabla_j\psi(\boldsymbol{\theta})$. This shows that the change in the image position is given by $\delta\theta_i = (A^{-1})_{ij}\,(\alpha_w)_j$, where $\boldsymbol{\alpha}_w=\boldsymbol{\nabla}_\theta\,t_w$ is the extra ray deflection due to weak lensing.

We are interested in the arrival time difference between the perturbed path and the unperturbed path. Since the value of $t(\theta)$ depends on the strong lens and is not known precisely, the effect is only observable as a time varying effect because the moving DM substructures transit the light paths over time. The arrival time difference is 
\begin{equation}
    t(\boldsymbol{\theta}^{'})-t(\boldsymbol{\theta})=\underbrace{\frac{1+z_d}{c}\frac{d_d\,d_s}{d_{ds}}\left(\frac{1}{2}(\boldsymbol{\theta^{'}+\theta})-\boldsymbol{\beta}-\boldsymbol{\nabla}_{\theta}\psi-\frac{1}{2}(\boldsymbol{\theta^{'}-\theta})\cdot\boldsymbol{\nabla}_{\theta}\boldsymbol{\nabla}_{\theta}\psi\right)\cdot(\boldsymbol{\theta^{'}-\theta})}_{\rm strong\, lensing}+\underbrace{(\boldsymbol{\theta^{'}-\theta})\cdot\boldsymbol{\nabla}_{\theta} t_w}_{\rm weak\,lensing}.
\end{equation}
Applying the original lens equation for the undisturbed path in Eq.~\ref{eq:lens_o}, the above equation can be simplified as
\begin{equation}
        t(\boldsymbol{\theta}^{'})-t(\boldsymbol{\theta})=\frac{1+z_d}{c}\frac{d_d\,d_s}{d_{ds}}\left(\frac{(\boldsymbol{\theta^{'}-\theta})^2}{2}-\frac{1}{2}\,\delta\theta_i\,\delta\theta_j\,\nabla_i\nabla_j\psi\right) +(\boldsymbol{\theta^{'}-\theta})\cdot\boldsymbol{\nabla}_{\theta}t_w,
\end{equation}
where $\delta\boldsymbol{\theta}=\boldsymbol{\theta}' - \boldsymbol{\theta}$. Now we can use Eq.~\ref{eq:dtheta} to determine the relation between the strong lensing term and weak lensing term 
\begin{equation}
    \begin{split}
        t(\boldsymbol{\theta}^{'})-t(\boldsymbol{\theta})
        &=\frac{1+z_d}{c}\frac{d_d d_s}{d_{ds}}\left(\frac{(\boldsymbol{\theta^{'}-\theta})^2}{2}-\frac{1}{2}\,\delta\theta_i\,\delta\theta_j\,\nabla_i\nabla_j\psi\underbrace{-(\boldsymbol{\theta^{'}-\theta})^2+\delta\theta_i\delta\theta_j\nabla_i\nabla_j\psi}_{\rm weak\,lensing}\right)\\
        &=\frac{1}{2}\boldsymbol{\nabla}_{\theta}t_w\cdot(\theta^{\prime}-\theta) = - \frac12\,\left( \frac{1+z_d}{c}\,\frac{d_d\,d_s}{d_{ds}}\right)^{-1}\,\boldsymbol{\nabla}_\theta t_w\cdot \boldsymbol{A}^{-1}\cdot \boldsymbol{\nabla}_\theta t_w.
    \end{split}
\end{equation}
The change in the arrival time $t(\boldsymbol{\theta}')-t(\boldsymbol{\theta})$ is therefore proportional to $(\alpha_w)_i\,(\alpha_w)_j\,(A^{-1})_{ij}$ and therefore is a correction of the second order in the weak lensing effect. It can also be seen that when the image has a large magnification factor, $\boldsymbol{A}$ is close to a singular matrix, and hence $t(\boldsymbol{\theta}')-t(\boldsymbol{\theta})$ is amplified by roughly the strong lensing magnification factor. Therefore, observable effects in the arrival time are greatly enhanced for a highly magnified FRB source, which leads to a better sensitivity to DM substructures.

The above result also enables us to directly apply previous calculations for the Shapiro time delay from DM substructures along two straight sightlines to the observable signal in the scenario of a multiply-imaged repeating FRB source. Note that over the time span of observations, which is usually on the order of years, the strongly lensed FRB source will remain strongly lensed since it will not move over a large distance on the length scale of galaxy lensing or galaxy cluster lensing.

For an example, we consider a strong lens profile described by the singular isothermal sphere (SIS), where $\psi \propto |\theta|$. The lens equation that determines $\delta\boldsymbol{\theta}$ can be decomposed into a perpendicular component and a parallel component with respect to $\beta$ in the 2D lens plane. The time delay fluctuation induced by line-of-sight DM substructures is
\begin{equation}
    \delta\theta_{\perp}\left(1-\frac{|\theta|}{|\theta-\beta|}\right) =-\nabla_{\perp}t_w, \qquad \delta\theta_{\parallel}=-\nabla_{\parallel}t_w,
\end{equation}
\begin{equation}\label{eq:delta_t_2i}
t(\boldsymbol{\theta}^{'})-t(\boldsymbol{\theta})   =-\frac{1}{2}\left(\frac{1+z_d}{c}\frac{d_d\,d_s}{d_{ds}}\right)^{-1}\left((\boldsymbol{\nabla}_\perp t_w)^2\left(1-\frac{|\theta|}{|\theta-\beta|}\right)^{-1}+(\boldsymbol{\nabla}_\parallel t_w)^2\right).
\end{equation}
The gradient of the weak lensing Shapiro time delay $t_w$ has already been calculated (Eq.~\ref{eq:sigmat}). The variance is simply given by the angular derivatives of the arrival time difference
\begin{equation}
    {\rm Var}(\nabla_\theta t_w) = \sigma_t^2D_s^2/x_{0}^2,
\end{equation}
When we measure the arrival time difference between two lensed images, the variance doubles. For Gaussian variables, $2({\rm Var}(\nabla_{\perp} t_w))^2={\rm Var}((\nabla_{\perp} t_w)^2)$. 
Therefore, the time variance of the two-image time delay can be estimated from Eq.~\ref{eq:delta_t_2i} as
\begin{equation}
        \sigma_{\rm 2i}^2\approx {\rm Var}\left(\frac{1}{2\,D_s}(\nabla_{\perp} t_w)^2\right)=\frac{1}{4}\frac{4\,\sigma_t^4\,D_s^2}{x_{0}^4}=\frac{\sigma_t^4\,D_s^2}{x_{0}^4}
\end{equation}
Therefore, we expect to see the arrival time variation with a variance 
\begin{equation}\label{eq:sigma_2i}
\begin{split}
        \sigma_{2i}&\approx \frac{4}{c^6}(4\pi G\bar{\rho}_{\rm m})^2D_s^2\int \frac{\rmd k_{\perp}}{2\pi}\frac{1}{k_{\perp}}P_{\delta}(k_{\perp},z) \,\left(1-e^{-k^2_{\perp}\sigma_v^2\delta t^2/2} \right)\\
        &\approx \frac{4}{c^6}(4\pi G\bar{\rho}_{\rm m})^2D_s^2\int \frac{\rmd k_{\perp}}{2\pi}\frac{1}{k_{\perp}}P_{\delta}(k_{\perp},z) \left( \frac{k^2_{\perp}\sigma_v^2\delta t^2}{2}-\frac{k^4_{\perp}\sigma_v^4\delta t^4}{8}+ ... \right)\\
\end{split}
\end{equation}
In the second line, the ellipsis represents higher-order terms we omit in the Taylor expansion of the exponential factor. However, the expansion will break down if $k\,\sigma_v\delta t \sim 1$ and we will have to use the temporal power spectrum defined in Eq.~\ref{eqn:Ptomega}.

The measurement of the time-varying time delay difference will be affected by the movement of the lens system, which is expected to cause the arrival time difference to change linearly over time~\cite{Dai2017lensedFRBtiming}. As we see in the above expression of the time delay, it will be useful to instead consider terms with higher-order temporal dependence for which mundane effects associated with long timescales are suppressed. The effect from the peculiar acceleration of the FRB source or that of the Earth is not significant enough to induce a time delay change over weeks. The geometric time delay caused by FRB accelerations will have a second derivative on time as $\sim x_0 \, \delta \ddot{x}_{\rm FRB}/D_s$, where $\delta x_{\rm FRB}$ is the change in the FRB source location. This rate is suppressed by the small ratio between the sightline separation and the source distance. For $\delta\ddot{x}_{\rm FRB}\sim 10^{-10}\rm m/s^2$, this corresponds to a time delay change of only $\sim 10^{-17}\,$s over a month.

Considering this secondary effect only, we can derive the constraints on the matter power spectrum as a function of the Fourier wavenumber $k$ by assuming that one logarithmic bin of $k$ dominates $P_\delta(k)$ at a time. The sensitivity on the matter power spectrum is given by 
\begin{equation}
    \delta P_{\delta}(k) \approx \frac{2\pi\,\delta t_m/\sqrt{N}}{\frac{2}{c^6}(4\pi G\,\bar{\rho}_{\rm m})^2\,D_s^2\,\sigma_v^2\,\delta t^2\,k^2},
\end{equation}
where $N$ is the number of well-timed radio bursts, $\delta t_m$ is the accuracy of the FRB timing, $D_s$ is the comoving distance of the FRB source, $\sigma_v$ is the characteristic velocity of dark matter structures, and $\delta t$ is the observation time. 
For the second-order effect discussed here, the sensitivity of a 2-Image system is equivalent to that of a 2-Dish system with a separation $x_0=N^{-1/4}\sqrt{2 D_s\,\delta t_m}\sim 5$ AU (The 2-Image system refers to strongly lensed FRBs with two images that provide two sightlines for free to the same FRB source). It is worth noting that we can not distinguish the second-order effect from the first-order effect and thus this only indicates the effect we calculate here is subdominant.

\section{Halo Model for the Nonlinear Power Spectrum}
\label{app:halo_model}

A challenge of making predictions for our FRB timing observable is that it probes nonlinear modes.  Enhancements in linear power can be destroyed by nonlinear evolution.  However, it has been shown that on small enough scales some enhancements are sufficient to survive \cite[e.g.][]{Xiao:2021nkb}, often as dense microhalos in larger halos.  Here we attempt to predict the power spectrum of such enhancements, with a focus on post-inflation axion and early matter domination models. 

To reliably predict if enhanced DM substructures can be probed with FRB timing, we calculate the nonlinear power spectrum using the halo model in place of the linear power spectrum. On scales smaller than the DM halo size, the one halo term should dominate the matter power spectrum, which is given by
\begin{equation}\label{eq:1_halo}
    P^{\rm 1h}(k)=\frac{1}{\bar{\rho}^2}\int \rmd M\,M^2\,\frac{\rmd n(M)}{\rmd M}\,|\tilde{u}(k|M)|^2
\end{equation}
where $\tilde{u}(k|M)$ is the Fourier transform of the mass density profile at a halo mass $M$, which we take to be the Navarro-Frenk-White (NFW) profile, and $n(M)$ is the halo mass function. The Fourier transform of the density profile can be calculated from the integral
\begin{equation}
    \tilde{u}(k|M)=4\pi\int_0^{\infty} u(r|M)\,\frac{{\rm sin} \, kr}{kr}\,r^2\,\rmd r,
\end{equation}
where $u(r|M)$ is the halo density profile divided by $M$. We would like to study NFW profiles that are truncated at the virial radius $r_{\rm vir}$, otherwise, the halo mass is logarithmically divergent. Therefore the density profile can be expressed as
\begin{equation}
    u(r|M) = \frac{1}{4\pi r\, r_s^2\,(1+r/r_s)^2}\,\left({\rm ln}(1+c)-\frac{c}{1+c} \right)^{-1},
\end{equation}
where $c=r_{\rm vir}/r_s$ is the halo concentration parameter and $r_s$ is the NFW scale radius. The median concentration parameter $c$ is a function of halo mass and redshift. For axion miniclusters, we use the mass-concentration relation in Ref.~\cite{Xiao:2021nkb}.

The halo mass function of the axion miniclusters, which form following initial white-noise density fluctuations, can be analytically calculated using the Press-Schechter formalism \cite{Dai:2019lud}, or using a modified Sheth-Tormen mass function calibrated by N-body simulations \cite{Xiao:2021nkb}. The overdensity variance of the relevant white-noise power spectrum is
\begin{equation}
   \sigma(M)=D_1(z)\,\left(\frac{3\pi A_{\rm osc}}{2}\frac{M_0}{M} \right)^{\frac12},
\end{equation}
where $D_1(z)$ is the growth function normalized at matter-radiation equality and $M_0=(4\pi/3)\,\bar{\rho}/k^3_{\rm osc}$ is the initial axion minicluster mass and $A_{\rm osc}$ is the amplitude of the axion perturbations at the horizon size.

Recent simulations of axion string network evolution suggest that $A_{\rm osc}\approx 0.03$, and that the matter power spectrum can be extended to subhorizon scales until the amplitude $\Delta(k)$ reaches order unity \cite{Vaquero:2018tib,Buschmann:2019icd}. This effect from string dynamics can be well captured by absorbing the departure of $A_{\rm osc}$ from unity through a redefinition of $k_{\rm osc}$, which results in a smaller effective axion minicluster mass $M_0$.
Note that we have only considered the white-noise power spectrum produced by the vacuum misalignment mechanism, which dominates the power spectrum on small scales. However, the population of axion miniclusters will eventually be affected by the formation of larger CDM halos, which will be discussed later.
The mass function of the axion miniclusters can be calculated using the Press-Schechter theory:
\begin{equation}
    \frac{\rmd n(M)}{\rmd M}= \sqrt{\frac{2}{\pi}}\frac{\bar{\rho}}{M^2}\left|\frac{\rmd {\rm ln}\sigma}{\rmd {\rm ln}\, M}\right|\frac{\delta_c}{\sigma(M)}\,\exp\left(-\frac{\delta_c^2}{2\,\sigma(M)^2}\right).
\end{equation}
Note that the one-halo term can reproduce the white-noise linear power spectrum in the low $k$ limit. In this case, $\tilde{u}(k|M)=1$ and the power spectrum is given by
\begin{equation}
     P^{\rm 1h}(k)|_{k\rightarrow 0}= \frac{1}{\bar{\rho}^2}\int \rmd M M^2\,n(M) =1.66\times\frac{4\pi}{3}\,\frac{D^2_1(z)}{k^3_{\rm osc}}.
\end{equation}
This has exactly the same scaling as the linear power spectrum expected from axion white-noise spectrum after applying the linear growth function, where $P_{\rm linear}=2\pi^2\,D_1(z)^2/k_{\rm osc}^3$. 
There is a mismatch in the normalization factor, which is expected since the one-halo term cannot account for the entirety of the density fluctuations at large scales. However, since the relevant scales for us are smaller than the sizes of the massive CDM halo, the one-halo term from axion miniclusters is the term that dominates the power spectrum on the small scales that contribute most to the signal. 

Now we are ready to use the white-noise-only mass function to obtain the full mass function of axion miniclusters and the corresponding nonlinear power spectrum.
Since axion miniclusters formed at earlier times will eventually be incorporated into massive CDM halos formed from adiabatic fluctuations, the full mass function in the late Universe should account for this effect. Here we assume that the axion miniclusters stop merging with each other once they fall into massive CDM halos as discussed in Ref.~\cite{Xiao:2021nkb}. Therefore the full mass function accounting for the incorporated halo is 
\begin{equation}
    \frac{\rmd n_{f}}{\rmd M}(z) = \left(1-f_{\rm col}^{\rm CDM}(z) \right) \frac{\rmd n_{\rm WN}}{\rmd M}(z) + \int_{z_{\rm eq}}^{z} \rmd z'\,\frac{\rmd f_{\rm col}^{\rm CDM}(z')}{\rmd z'}\,\frac{\rmd n_{\rm WN}}{\rmd M}(z'), 
\end{equation}
where $\rmd n_{\rm WN}/\rmd M$ is the white-noise-only mass function of axion miniclusters and $f_{\rm col}^{\rm CDM}$ is the collapse fraction of massive CDM halos, which is calculated according to the adiabatic initial density fluctuations from inflation. This mass function accounts for axion miniclusters absorbed by larger CDM halos at different redshifts. It provides a good benchmark model for the realistic nonlinear matter power spectrum of axion miniclusters. We use the Press-Schechter theory to calculate the collapse fraction of massive CDM halos, which gives
\begin{equation}
    f_{\rm col}^{\rm CDM}(z)={\rm erfc}\left(\frac{\delta_c}{\sqrt{2}\,\sigma_{\rm CDM}(M_{\rm min})\,D(z)}\right),
    \label{eqn:fcol}
\end{equation}
where $\sigma_{\rm CDM}(M)$ is the RMS of the filtered initial overdensity field without including the contribution of the white-noise spectrum from axion string network evolution, and $M_{\rm min}$ corresponds to the least massive CDM halo that can host axion miniclusters, which is set to $10^{-2}\,M_{\odot}$ in the calculation. Since $\sigma_{\rm CDM}(M)$ only depends on $M$ logarithmically, the result is not very sensitive to our choice of $M_{\rm min}$. Inserting this mass function back in Eq.~\ref{eq:1_halo}, we can calculate the nonlinear matter power spectrum induced by axion miniclusters on very small scales.

\begin{figure}[h!]
    \centering
\includegraphics[width=12cm]{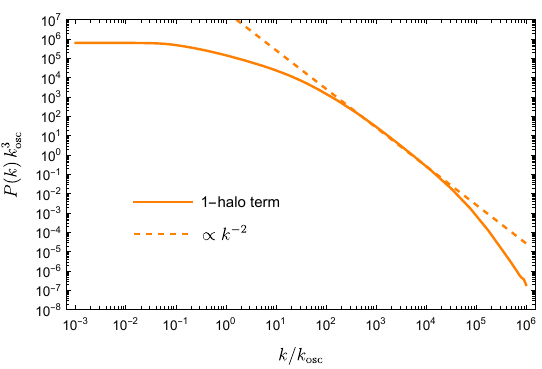}
\caption{The nonlinear matter power spectrum due to axion miniclusters, calculated from the one-halo term in the halo model. The results are plotted in terms of the dimensionless variables $k/k_{\rm osc}$ and $k_{\rm osc}^3\,P(k)$, so that the results shown here can be applied to all non-linear matter power spectrum arising from an initial white-noise power spectrum of an arbitrary amplitude in the same cosmology. At small $k$ values, the nonlinear power spectrum approaches a constant and behaves in the same way as a white-noise power spectrum. At high $k$ values, the nonlinear power spectrum decreases faster than $k^{-2}$.
}
    \label{fig:power}
\end{figure}

The non-linear matter power spectrum is shown in Fig.~\ref{fig:power}. The non-linear matter power spectrum is slightly suppressed than the linear extrapolation at $k\lesssim k_{\rm osc}$ but can extend to $k> k_{\rm osc}$ since the internal density profiles of axion miniclusters introduce power on small scales.
Since $1/k_{\rm osc}$ is roughly the size of the smallest axion minicluster formed from the white-noise linear power spectrum, it is not surprising that the true power spectrum is extended to $k>k_{\rm osc}$ once the miniclusters form. However, there might be a physical cutoff in the non-linear power spectrum, which is related to the formation of axion stars \cite{Eggemeier:2019jsu,Fox:2023aat}. A peak of sufficiently large density will lead to the formation of Bose-Einstein condensate states called axion stars, which will be unstable to axion self-interaction if the density exceeds some critical value. In this work, we conservatively take $10^4\,k_{\rm osc}$ to be a fiducial cutoff on the non-linear power spectrum, with the caveat that there could be interesting additional power at higher $k$ values that our observable could probe. However, the nonlinear power spectrum decreases much faster at higher $k$, as shown in Fig.~\ref{fig:power}, which means the effects may be difficult to detect. The decreasing matter power spectrum at high $k$ is caused by the mass density profile, whose Fourier transform $\tilde{u}(k|M)$ saturates in the low $k$ limit.


\begin{figure}[h!]
    \centering
\includegraphics[width=12cm]{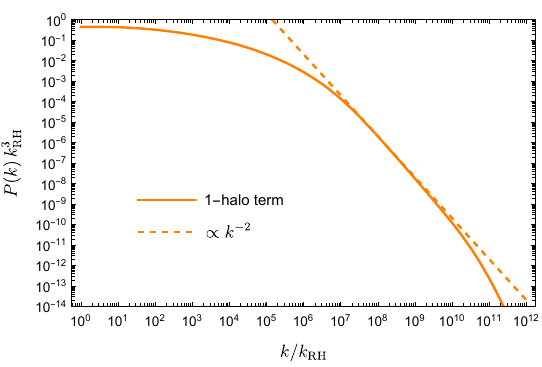}
\caption{The nonlinear matter power spectrum due to DM minihalos produced in a scenario of early matter domination, calculated from the one-halo term in the halo model. The results are plotted in terms of the dimensionless variables $k/k_{\rm RH}$ and $k_{\rm RH}^3\,P(k)$, with $k_{\rm RH}$ given by Eq.~\ref{eq:kRH}. At small $k$ values, the nonlinear power spectrum approaches a constant and behaves as a white-noise power spectrum. At high $k$ values, the nonlinear power spectrum decreases faster than $k^{-2}$, which is qualitatively similar to the case of axion miniclusters. }
\label{fig:power_emd}
\end{figure}

The same calculation can be applied to the scenario of early matter domination \cite{Erickcek:2011us}. The characteristic length scale of DM minihalo collapse produced during an early epoch of matter domination is related to the comoving horizon size when the Universe is at the reheating temperature $T_{\rm RH}$:
\begin{equation}\label{eq:kRH}
    k_{\rm RH} = 0.01\,{\rm pc^{-1}}\,\left(\frac{T_{\rm RH}}{1\,\rm MeV }\right)\left(\frac{10.75}{g_{* S}(T_{\rm RH})}\right)^{1/3}\left(\frac{g_{*}(T_{\rm RH})}{10.75}\right)^{1/2}.
\end{equation}
This corresponds to a characteristic DM minihalo mass $M_{\rm RH} = (4\pi/3)\,\bar{\rho}/{k_{\rm RH}^3}$. The RMS of the linear matter overdensity fluctuations, extrapolated to the present epoch and filtered over a length scale corresponding to the collapse of minihalos of mass $M$, can be expressed as $\sigma(M\lesssim M_{\rm RH})=2.3\,(M/M_{\rm RH})^{-0.66}$.

The nonlinear matter power spectrum in the scenario of early matter domination is shown in Fig.~\ref{fig:power_emd}. The shape of the nonlinear power spectrum is largely similar to that in the scenario oif axion miniclusters, except for the overall amplitude at the characteristic comoving wave number, $k_{\rm osc}$ or $k_{\rm RH}$, which corresponds to the comoving horizon size at the epoch of axion oscillation or at reheating, respectively. The amplitude difference is understandable, since the axion miniclusters collapse from large inhomogeneity sourced by axion isocurvature perturbations, while early matter domination only causes an extra amount of growth for the inflationary adiabatic initial fluctuations.

\section{Supernovae Lensing}
\label{app:sn_lensing}

The variance of the distance modulus of Type Ia supernovae is affected by weak lensing of intervening matter and given by \citep[e.g.][]{2006PhRvD..74f3515D}
\begin{equation}\label{eq:sigma_cos}
    \sigma^2_{\rm cos}=\frac{225\pi\,\Omega_m^2\,H_0^4}{4\,[{\rm ln(10)}]^2}\int_0^{\chi_s}\rmd \chi\,[1+z(\chi)]^2\,\frac{\chi^2\,(\chi_s-\chi)^2}{\chi_s^2}\int_0^{\infty}\frac{\rmd k}{2 \pi^2 }k\,P_\delta(k,z(\chi)),
\end{equation}
where $\chi$ is the comoving distance to the source. The $k$ weighting in the integrand of the $k$ integration is the same as in the linear term in Eq.~\ref{eqn:sigmat}. Type Ia supernovae are standard candles so any variance induced externally by weak lensing is constrained to be less than how well their luminosities can be standardized. In particular, present observations of Type Ia supernovae constrains any additional dispersion to $\sigma^2_{\rm cos}<(0.1)^2$ \citep{2006PhRvD..74f3515D}. The weak lensing contribution from the standard $\Lambda$CDM matter power spectrum is almost detectable as $\sigma_{\rm CDM}^2=(0.05)^2$ \citep{2006PhRvD..74f3515D}. Any additional weak lensing contribution from matter lumpiness on very small scales could be ruled out by such observations.

Assuming supernovae are roughly located at $z=1$, we can translate the bound on $\sigma^2_{\rm cos}$ to a constraint on the dimensionless matter power spectrum in a logarithmic bin:
\begin{equation}
    \frac{\delta P_\delta(k) k^3}{2\pi^2}\approx 4.4\times 10^5\left(\frac{k}{\rm kpc^{-1}}\right).
    \label{eqn:sn}
\end{equation}
Compared to the sensitivity of the 2-Dish configuration as shown in Eq.~\ref{eq:Pk_t1es}, and assuming $0.1\,$ns timing, $\delta t= 10\,$yr, $D_s = 1\,$Gpc and $N=100$, the constraint given by Eq.~\ref{eqn:sn} is comparable to our sensitivity to a linear time variation in the FRB arrival time difference for a spatial baseline of 0.14 AU. The supernovae constraint is valid only on length scales larger than that of the supernova photosphere weeks after the explosion, which corresponds to $k \lesssim 10^4$\,pc$^{-1}$. A related analysis of the constraints on the matter power spectrum from weak lensing of supernovae was done in Ref.~\cite{Ben-Dayan:2015zha}, with a focus on power enhancments at lower wavenumbers.  We focus on higher wavenumbers where enhanced nonlinear power can survive via dense microhalos.

\bibliographystyle{apsrev4-2}
\bibliography{FRB_NP}
\end{document}